\documentclass[%
 reprint,
superscriptaddress,
 amsmath,amssymb,
 aps,pra
]{revtex4-2}
\usepackage{siunitx}
\usepackage{xcolor}
\usepackage{graphicx}
\usepackage{dcolumn}
\usepackage{bm}
\usepackage{xspace}
\usepackage{xpatch}
\usepackage{amsmath}
\usepackage{tikz}
\usepackage{hyperref}

\def\Hbarplus   {\ensuremath{\overline {\mathrm{H}}^+}\xspace}
\def\Hbar   {\ensuremath{\overline {\mathrm{H}}}\xspace}
\def\gbar   {\ensuremath{\overline {g}}\xspace}
\def\bR     {\ensuremath{\mathbf{R}}\xspace}
\def\cL     {\ensuremath{\mathcal{L}}\xspace}
\def\cI     {\ensuremath{\mathcal{I}}\xspace}
\def\md     {\ensuremath{\mathrm{d}}\xspace}
\def\mc     {\ensuremath{\mathrm{c}}\xspace}
\def\mf     {\ensuremath{\mathrm{f}}\xspace}

\begin{document}
\title{Improving the statistical analysis of anti-hydrogen free fall by using near edge events}

\author{Olivier Rousselle}
\affiliation{%
 Laboratoire Kastler Brossel, Sorbonne Universit\'e, CNRS, ENS-PSL, Coll\`ege de France, 4 place Jussieu, 75005 Paris, France
 }%
\author{Pierre Clad\'e}%
\affiliation{%
 Laboratoire Kastler Brossel, Sorbonne Universit\'e, CNRS, ENS-PSL, Coll\`ege de France, 4 place Jussieu, 75005 Paris, France 
}%
\author{Sa\"ida Guellati-Kh\'elifa}%
\affiliation{%
 Laboratoire Kastler Brossel, Sorbonne Universit\'e, CNRS, ENS-PSL, Coll\`ege de France, 4 place Jussieu, 75005 Paris, France 
}\affiliation{Conservatoire National des Arts et M\'etiers, 292 rue Saint Martin, 75003 Paris, France
}%
\author{Romain Gu\'erout}
\affiliation{%
 Laboratoire Kastler Brossel, Sorbonne Universit\'e, CNRS, ENS-PSL, Coll\`ege de France, 4 place Jussieu, 75005 Paris, France 
}%
\author{Serge Reynaud}
\affiliation{%
 Laboratoire Kastler Brossel, Sorbonne Universit\'e, CNRS, ENS-PSL, Coll\`ege de France, 4 place Jussieu, 75005 Paris, France 
}%


\date{\today}

\begin{abstract}
An accurate evaluation of the gravity acceleration from the timing of free fall of anti-hydrogen atoms in the GBAR experiment requires to account for obstacles surrounding the anti-matter source. These obstacles reduce the number of useful events but may improve accuracy since the edges of the shadow of obstacles on the detection chamber depends on gravity, bringing additional information on the value of $g$. We perform Monte Carlo simulations to obtain the dispersion and give a qualitative understanding of the results by analysing the statistics of events close to an edge. We also study the effect of specular quantum reflections of anti-hydrogen on surfaces and show that they do not degrade the accuracy that much.
\end{abstract}

                             
\maketitle

\section{Introduction}
\label{sec:introduction}

One of the fascinating questions which remain open in modern physics is the asymmetry between matter and antimatter observed in the Universe but not fully accounted for in the Standard Model 
\cite{Hori2013,Bertsche2015,Charlton2017,Yamazaki2020}. 
In particular experimental tests of the effect of gravity on antimatter must still be improved \cite{Alpha2013}.
Ambitious projects are currently developed at new CERN facilities to produce low energy anti-hydrogen ($\Hbar$) atoms \cite{Maury2014} and measure $\gbar$, the gravity acceleration of neutral $\Hbar$ atoms \cite{Bertsche2018,Pagano2020,Mansoulie2019}. Among these projects, the GBAR experiment (\emph{Gravitational Behaviour of Anti-hydrogen at Rest}) aims at timing the free fall of ultra-cold $\Hbar$ atoms \cite{Indelicato2014,Perez2015}. 
Knowing the sign and order of magnitude of $\gbar$ would already be an important achievement, and improving the accuracy of its measurement would be crucial for advanced tests of the Equivalence Principle in the line of the many high precision tests performed on matter objects \cite{Wagner2012,Touboul2017,Will2018,Viswanathan2018,Asenbaum2020}.

The principle of the GBAR experiment is based upon an original idea of Hänsch and Walz \cite{Walz2004}. Anti-hydrogen ions $\Hbarplus$  are cooled in an ion trap by using laser cooling techniques.  The excess positron is photo-detached with a laser, forming a neutral anti-hydrogen atom $\Hbar$, with the laser pulse marking the start of the free fall. The end of free fall is timed by the annihilation of $\Hbar$ on the detection surface and the acceleration $\gbar$ is deduced from a statistical analysis of annihilation events. In a previous work \cite{Rousselle2021}, we have analysed the accuracy on $\gbar$ to be expected in a simple geometry for the GBAR experiment, taking into account the impact of the photo-detachment process on the initial velocity distribution and the statistics of annihilation events. 
We also noticed that the accuracy could be improved by considering the ceiling that intercepts some of the trajectories.

In the present paper, we go further in this analysis by taking into account the obstacles surrounding the anti hydrogen source, required for the experiment \cite{Hilico2014,Sillitoe2017}. These obstacles, such as the electrodes of the ion trap, intercept some trajectories of $\Hbar$ atoms. As for the ceiling in \cite{Rousselle2021}, one might think that they degrade the accuracy as they reduce the number of annihilation events used for the measurement. We show that the opposite happens, with an accuracy on the measurement of $\gbar$ improved thanks to the additional information gained from events close to the edges of the shadow of obstacles. 

We will first specify the geometry (\S\ref{sec:geometry}) and give a detailed simulation of annihilation events in the presence of obstacles to calculate the dispersion of the free fall measurement (\S\ref{sec:dispersion}). We will then give a qualitative understanding of the results by analysing the statistics of events close to an edge of the shadow of obstacles (\S\ref{sec:events}). 
We will finally make the analysis more complete by evaluating the effect of quantum reflection of $\Hbar$ atoms on the Casimir-Polder potential in the vicinity of matter surfaces \cite{Dufour2013}. Including this effect in the statistical analysis of the experiment, we will show that quantum reflection does not degrade the accuracy that much (\S\ref{sec:effect}). We assume quantum reflection to be specular, which requires the surfaces exposed to anti-atoms to be well polished.

\section{Geometry of the experiment}
\label{sec:geometry}

\begin{figure}[t!]
\includegraphics[scale=0.7]{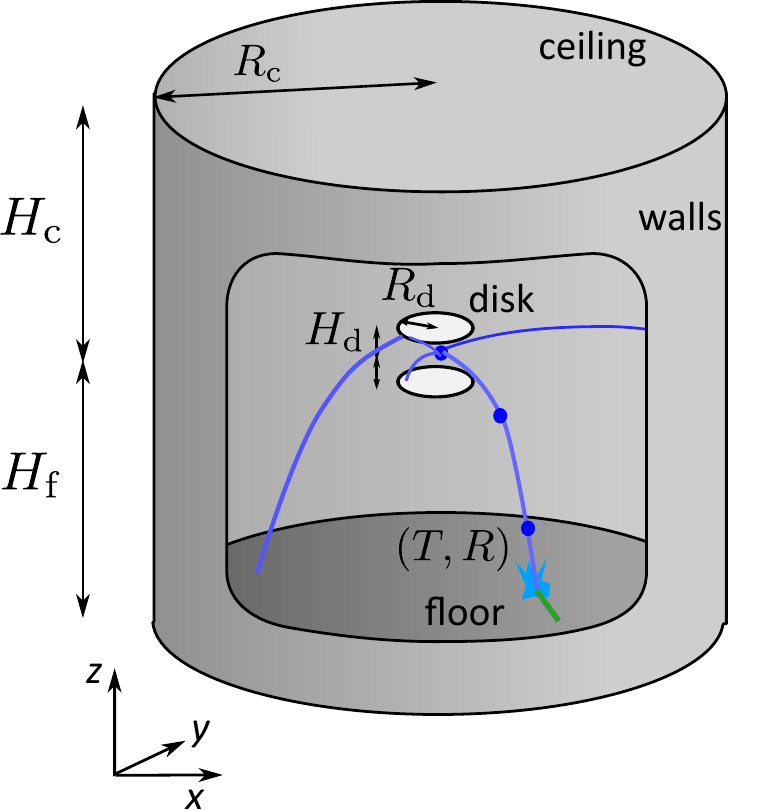}
\caption{\label{fig:trap}  Schematic representation of the GBAR free fall chamber with two disks symmetrically positioned above and below the trap to mask the obstacles surrounding it. Trajectories to the surfaces of the chamber are represented as blue lines. }
\end{figure}

The source of $\Hbar$ atoms is placed at the centre of the cylindrical vacuum chamber (radius $R_\mc$ and free fall height $H_\mf$) in which the free fall measurement is performed. This source is surrounded by obstacles such as the electrodes of the trap \cite{Hilico2014,Sillitoe2017}.
We define a cleaner geometry by hiding obstacles with two symmetrically positioned disks of radius $R_\md$ placed above and below the trap at a distance $H_\md$. 
The resulting geometry is shown schematically on Fig.\ref{fig:trap}, with trajectories to the surfaces of the chamber represented as blue lines \cite{Dufour2013}.

The symmetrical configuration produces a simple geometry which will be more easily studied in Monte-Carlo simulations of the experiment. We will work with a horizontal polarisation of the photo-detachment laser, in order to launch the atoms preferably in the free interval between the two disks \cite{Rousselle2021}. In a first part of the study we will generate random events mimicking the forthcoming experiment with a reference value $g_0=9,81$ m/s$^2$. In a second part, we will present the statistical analysis of these events mimicking the data analysis process to be developed at a later stage for the experiment. 
The whole analysis will be done in a manner quite analogous to that presented in \cite{Rousselle2021}, with however important differences discussed now.  

The evaluation of $g$ from the analysis of annihilation data involves the calculation of the probability current $J(\bR,T)$ (number per unit of surface and unit of time) to detect a particle at position $\bR$ in space and $T$ in time. In \cite{Rousselle2021}, we detailed the calculation of the same quantity $J_0(\bR,T)$ ignoring the presence of obstacles.
In the presence of symmetrical disks in contrast, the reasoning has to consider separately the annihilation events on the surfaces of the free fall chamber, which are used for estimating $g$, and those on the disks, which contain essentially no information on $g$. 
Hence, we will be mainly interested in the current on the surfaces of the free fall chamber with a probability integral $P_\mc$ smaller than one. 
In the following we fix the initial number $N$ of atoms but our analysis of dispersion accounts for the fact that the number of events $N_\mc=NP_\mc$ detected on the surfaces of the chamber is smaller than $N$.  

\begin{figure}[t!]
\includegraphics[width=\linewidth]{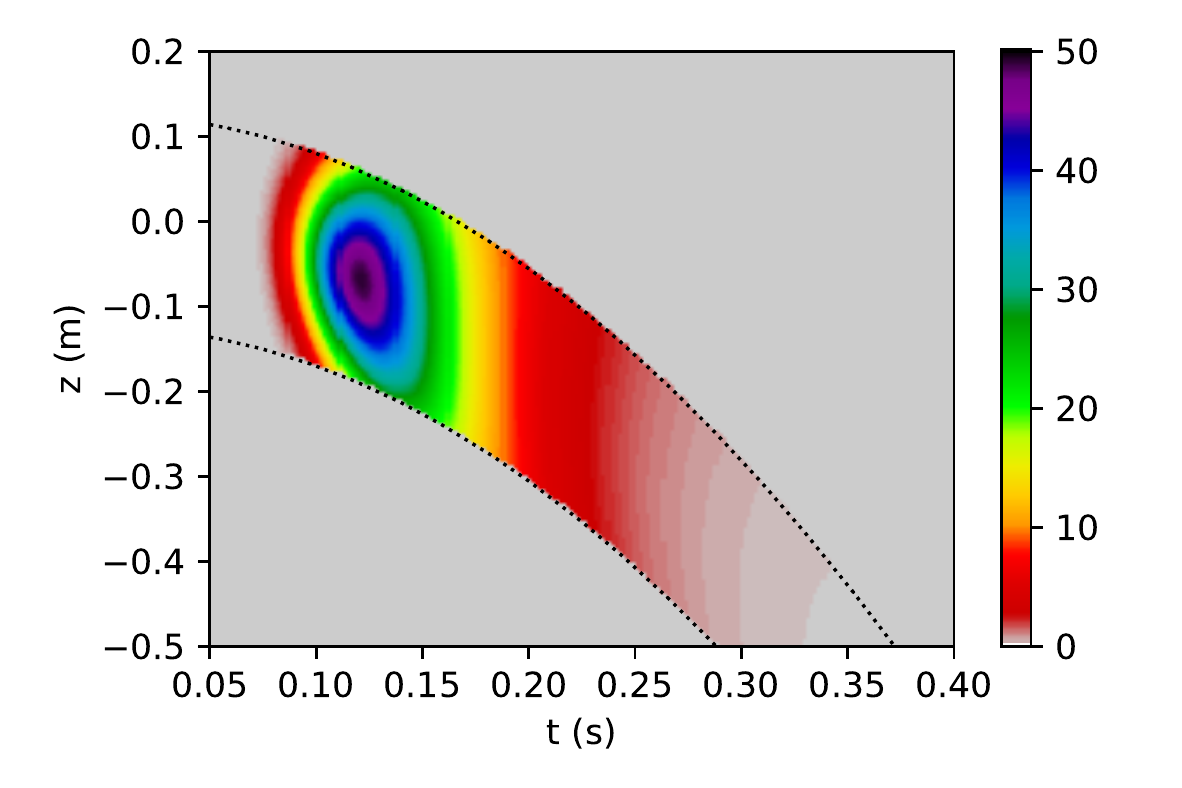}
\caption{Distribution of annihilation events ($j(\bR,t)$ in \si{s^{-1}m^{-2}}) on the wall as a function of $t$ and position ($X=0, Y=R, Z$) with two disks. Parameters are $R_\md=\SI{2}{\centi\meter}$, $H_\md=\SI{1}{\centi\meter}$, $R_\mc=\SI{25}{\centi\meter}$, $f=\SI{1}{\mega\hertz}$ and  $\delta E=\SI{30}{\micro\electronvolt}$.}
\label{fig:figure_J_z_T_two_diska}
\end{figure}

At the end of calculations, we will obtain the mean $\mu_g$ and the standard deviation $\sigma_g$ of the estimator defined for $\gbar$, simply denoted $g$ from now on.  
In spite of the loss of useful events, it will turn out that the standard deviation $\sigma_g$ may be smaller in the presence of the obstacles.
The main reason for this important result can already be understood by looking at the detection current on the walls of the free fall chamber represented on Fig.\ref{fig:figure_J_z_T_two_diska}.
One clearly sees on this figure the sharp boundaries of the shadow induced on the walls by the presence of the disks. The position in space and time of this shadow depends on the value of $g$ and its detection allows to gain information on the value of $g$.

\begin{figure}[b!]
\includegraphics[width=\linewidth]{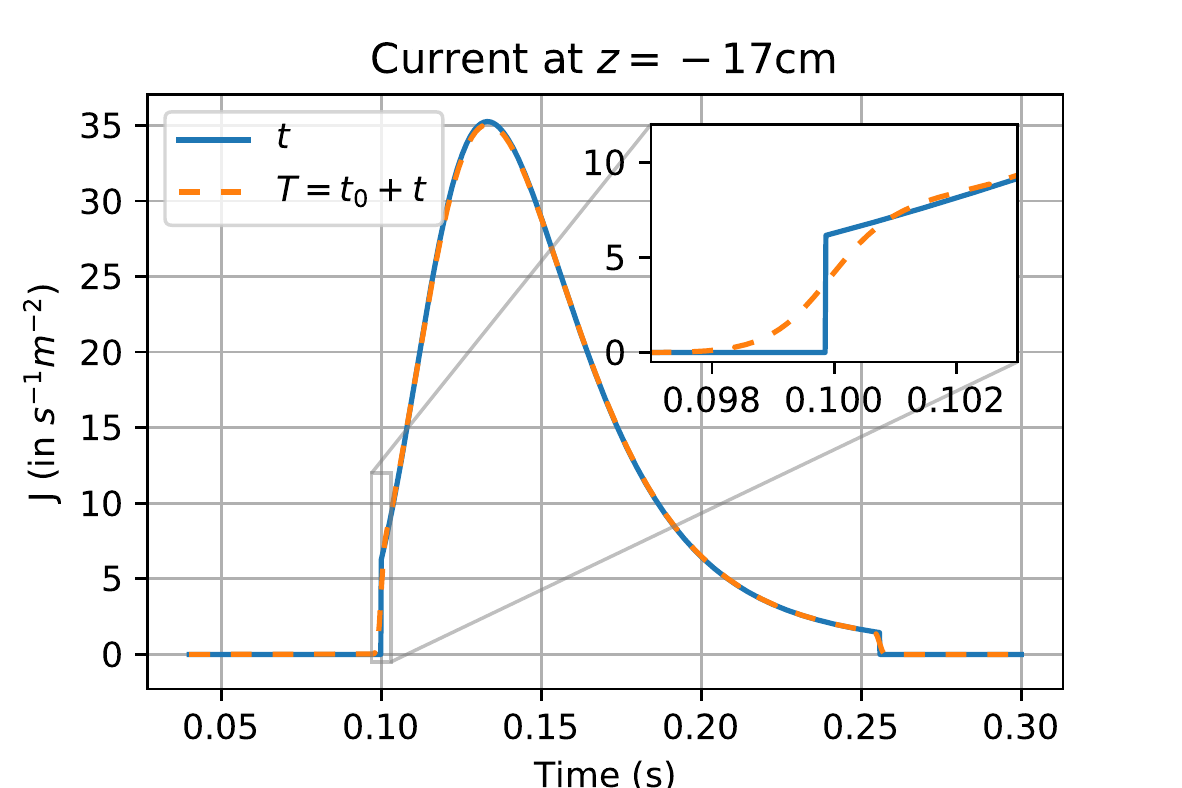}
\caption{Comparison between the currents $j(\bR,t)$ and $J(\bR,T)$ calculated with the same parameters as in Fig.\ref{fig:figure_J_z_T_two_diska}, the first one before the convolution, the second one after the convolution with $\tau=\SI{500}{\micro\second}$. The effect of the dispersion on $t_0$ is visible on the edges of the shadow zone induced by the obstacles.}
\label{fig:figure_J_z_T_two_diskb}
\end{figure}

\begin{figure*}[t!]
\includegraphics[width=0.66\linewidth]{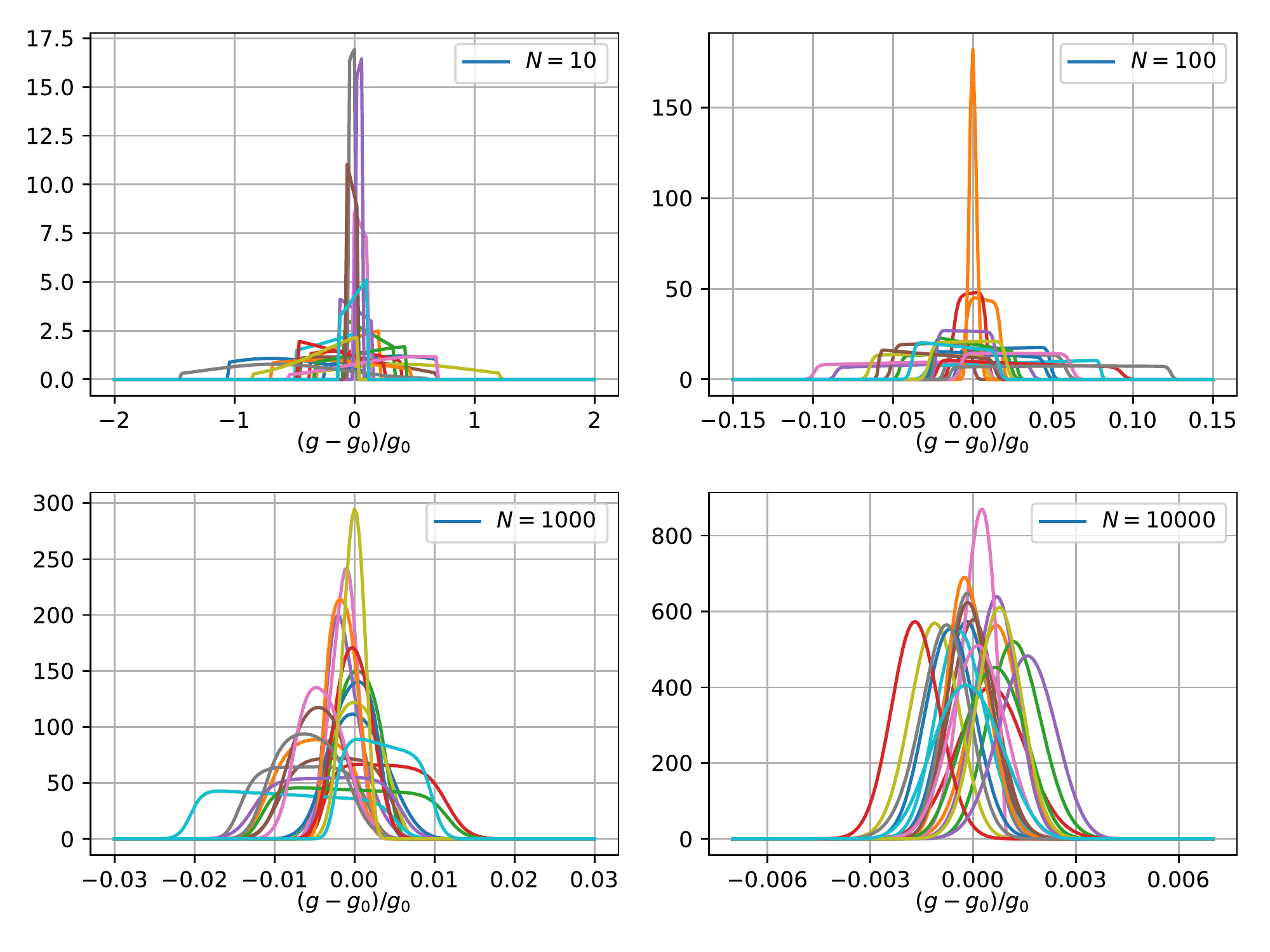} 
\caption{Set of normalised likelihood $\ell(g)$ for different values of the initial number of atoms $N$. Same parameters as for Fig.\ref{fig:figure_J_z_T_two_diskb}.}
\label{fig:normalised_likelihood_smoothing}
\end{figure*}

The dispersion of initial position of the ion in the trap plays a negligible role in the problem, while the dispersion of the photo-detachment time $t_0$ has to be accounted for as in \cite{Rousselle2021}. The time of annihilation event $T$ is $t+t_0$, where $t$ is the time of flight and $t_0$ the precise time of the photo-detachment event. Hence the current $J(\bR,T)$ (taking into account the dispersion on $t_0$) is calculated as the convolution of a current $j(\bR,t)$ neglecting this dispersion and the distribution of $t_0$, assumed to be a logistic distribution with width $\tau$
\begin{eqnarray}
&&J(\bR,T)=\int j(\bR,T-t_0) \, \delta_\tau(t_0) \, \md t_0
~,\notag \\
&&\delta_\tau(t_0) = 
\frac{1}{4\tau}\frac1{\cosh^2\left(\frac {t_0}{2\tau}\right)} ~.
\label{eq:logistic}
\end{eqnarray}

The current plotted in Fig.\ref{fig:figure_J_z_T_two_diska}
has been calculated before the convolution, and the latter will round up the edges of the shadow zone without suppressing the gain of information associated with them.
Currents calculated before and after the convolution on a cut with fixed altitude ($z=-0.17$cm) are represented on  Fig.~\ref{fig:figure_J_z_T_two_diskb}. The effect of the dispersion on $t_0$, calculated here for $\tau=\SI{500}{\micro\second}$, is visible at the steps of the current corresponding to edges of the shadow of the disks. We will see that it plays an important role in some forthcoming calculations while it can be neglected in other ones.

 \section{Dispersion of the estimator}
\label{sec:dispersion}

In this section, we present Monte-Carlo simulations to discuss the dispersion to be expected on the measurement of $g$ with the obstacles taken into account. We go rapidly on steps which were already discussed in \cite{Rousselle2021} for the case without obstacles and discuss mainly the differences with this case. 

 Considering a draw of $N$ $\Hbar$ atoms that escape from the trap after the photo-detachment process, we calculate the trajectory that depends on the random initial velocity $\bm{v}_0$ and the random time of photo-detachment $t_0$ and deduce the annihilation position in space $\bR_i$ and time $T_i$. Trajectories hitting the disk lead to annihilation there, and are discarded from the forthcoming analysis as they contain no useful information on the value of $g$. 
 
 We thus calculate the likelihood function $\cL$ and normalised likelihood function $\ell$ for the draw of $N_\mc$ events that annihilate on the surfaces of the chamber
 \begin{equation}
\cL (g) = \prod_{i=1}^{N_\mc} J({\bR _i}, t_i)
\quad,\quad
\ell (g) = \frac{\cL (g)}{\int \cL (g) \md g} ~.
\label{eq:def_likelihood}
\end{equation}
 Normalised likelihood functions $\ell (g)$ are represented on Fig.~\ref{fig:normalised_likelihood_smoothing} for a given set of parameters ($R_\md=\SI{2}{\centi\meter}$, $H_\md=\SI{1}{\centi\meter}$, $R_\mc=\SI{25}{\centi\meter}$, $f=\SI{1}{\mega\hertz}$, $\delta E=\SI{30}{\micro\electronvolt}$) and four values of $N$.
 The different functions $\ell (g)$ plotted for each case are calculated for independent random draws.
 
 For $N=10$ and $N=100$, the likelihoods are mostly flat with sudden drops to zero. This behaviour is due to the obstacles and can be qualitatively understood with $\tau=0$. Let us consider an impact at $\bR,t$ reached by an atom for $g=g_0$. If this impact is close to an edge of the allowed area, it may fall in the shadow zone for a different value $g\neq g_0$, so that the likelihood drops to zero. The drop to zero is rounded up by the dispersion $\tau$, with the rounding negligible for $N=10$ or $N=100$ but starting to be noticeable for $N=1000$. For $N=10000$, the likelihoods are closer to Gaussian functions because the numerous annihilation events produce an efficient sampling of the rounded step.

Though all likelihood functions are centred around the expected value $g_0$, their maximum will fall on either side of their plateau so that the common maximum likelihood estimator will show large variations. In order to circumvent this problem, we define another estimator $\check g$ as the mean value of the likelihood $\ell$ and will use it in all forthcoming simulations
\begin{equation}
\check g = \int g \,\ell(g) \md g = \frac{\int g\,\cL (g) \md g}{\int \cL (g) \md g}   ~.
\label{MeanEstimator}
\end{equation}

\begin{figure}[t!]
\centering\includegraphics[width=\linewidth]{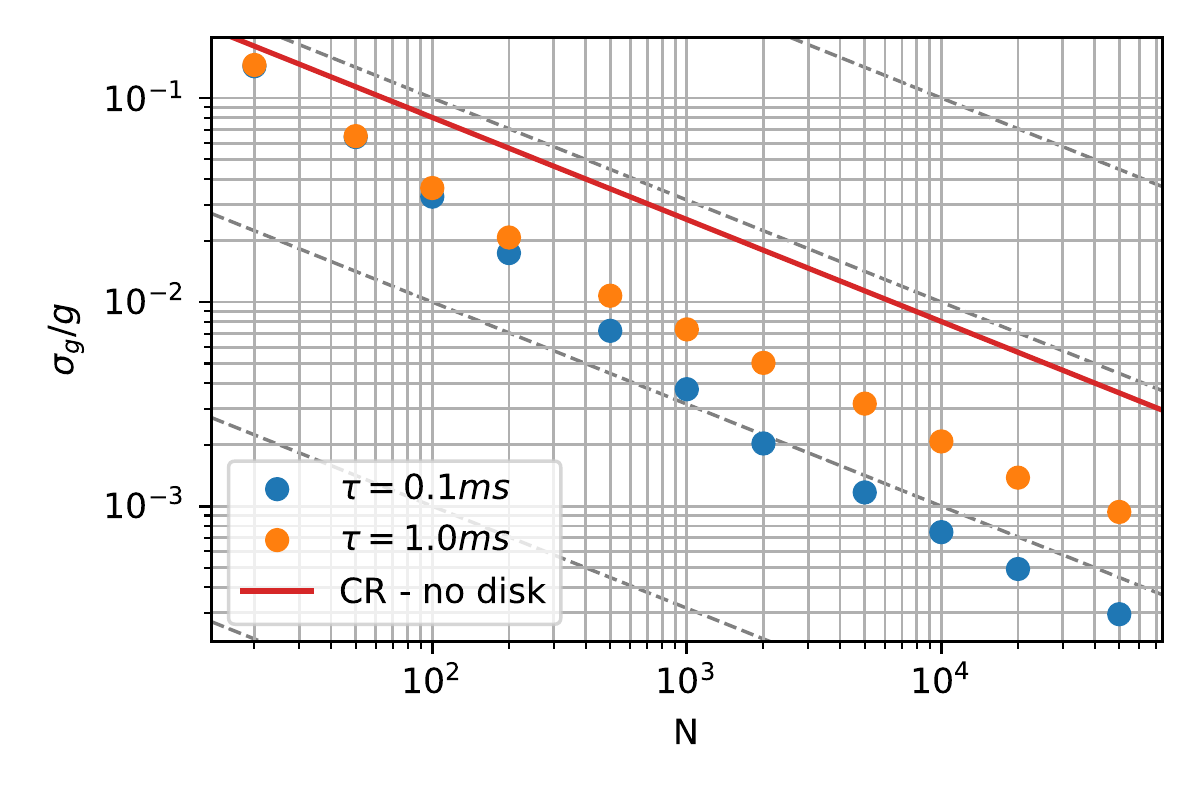}
\caption{Relative dispersion obtained in Monte-Carlo simulation of the experiment with two disks. Orange and blue dots represent respectively the dispersion for $\tau=\SI{1}{ms}$ and $\tau=\SI{0.1}{ms}$. The red line is the Cramer-Rao limit without disks. The light grey lines are guides for the eye showing a $1/\sqrt{N}$ scaling.}
\label{fig:disk_simulation_tau}
\end{figure}

Fig.~\ref{fig:disk_simulation_tau} shows the relative dispersion of $\check g$ as a function of the number $N$ of initial events. The dispersion of $t_0$ is taken into account for all calculations, though it has a small effect for small value of $N$. We see that the variation of the dispersion versus $N$ does not follow a $1/\sqrt{N}$ except for very large values of $N$. This behaviour is an indication that the statistical efficiency is reached only for those very large values. In high $N$ regime the dispersion depends on $\tau$ and it is smaller than the dispersion without the obstacles.

\section{Statistics of events close to an edge}
\label{sec:events}

We now present two methods which are useful to understand the results of the simulations in the two regimes discussed at the end of \S\ref{sec:dispersion}.
These two methods deal with the statistics of events close to an edge, first in the sharp case ($\tau=0$) better suited to small values of $N$, then in the rounded case ($\tau\neq0$) better suited to large $N$.

\subsection{The min-max model}
\label{subsec:theminmaxmodel}

We first discuss the drop in the likelihood observed for values of $N$ such as $N=100$. The sampling of the edges of the shadow zone is not efficient in this case so that we can neglect the dispersion $\tau$ of $t_0$ and simplify calculations by using the current $j(\bR,t)$ before convolution (so that $T\equiv t$). 

For a given impact $\bR,t$, we calculate the initial velocity assuming a value of $g$. We define a function $\lambda_g(\bR,t)$ equal to 1 if the associated trajectory reaches the detection point without hitting the disks, and equal to 0 in the opposite case where $\Hbar$ is annihilated on the disks. We get the current $j$ as the product of this function by the current $j_{0}$ calculated without obstacles
\begin{equation}
j(\bR,t) = \lambda_g(\bR , t) j_{0}(\bR , t) ~.
\label{eq:J_obs}
\end{equation}

For a random draw of $N$ atoms, the likelihood function \eqref{eq:def_likelihood} is then written as
\begin{eqnarray}
&&\cL (g) = \cL _0(g) \pi(g) ~,\quad
\cL _0(g) = \prod_{i=1}^{N} j_0({\bR _i}, t_i)
~,\notag\\
&&\pi(g) = \prod_{i=1}^{N} \lambda_g(\bR _i, t_i) ~,
\end{eqnarray}
with $\cL _0(g)$ the likelihood function calculated without obstacles. Meanwhile $\pi(g)$ is a rectangular function, with a unit value on the interval between a minimal and maximal value of $g$ which are random variables depending on the full set of impact parameters for the $N_\mc$ events
\begin{equation}
\pi(g) = \begin{cases}
1 & \mathrm{if~} g\in[g_{\min} , g_{\max} ] \\
0 & \mathrm{otherwise}
\end{cases}~.
\end{equation}

\begin{figure}[t!]
\includegraphics[width=\linewidth]{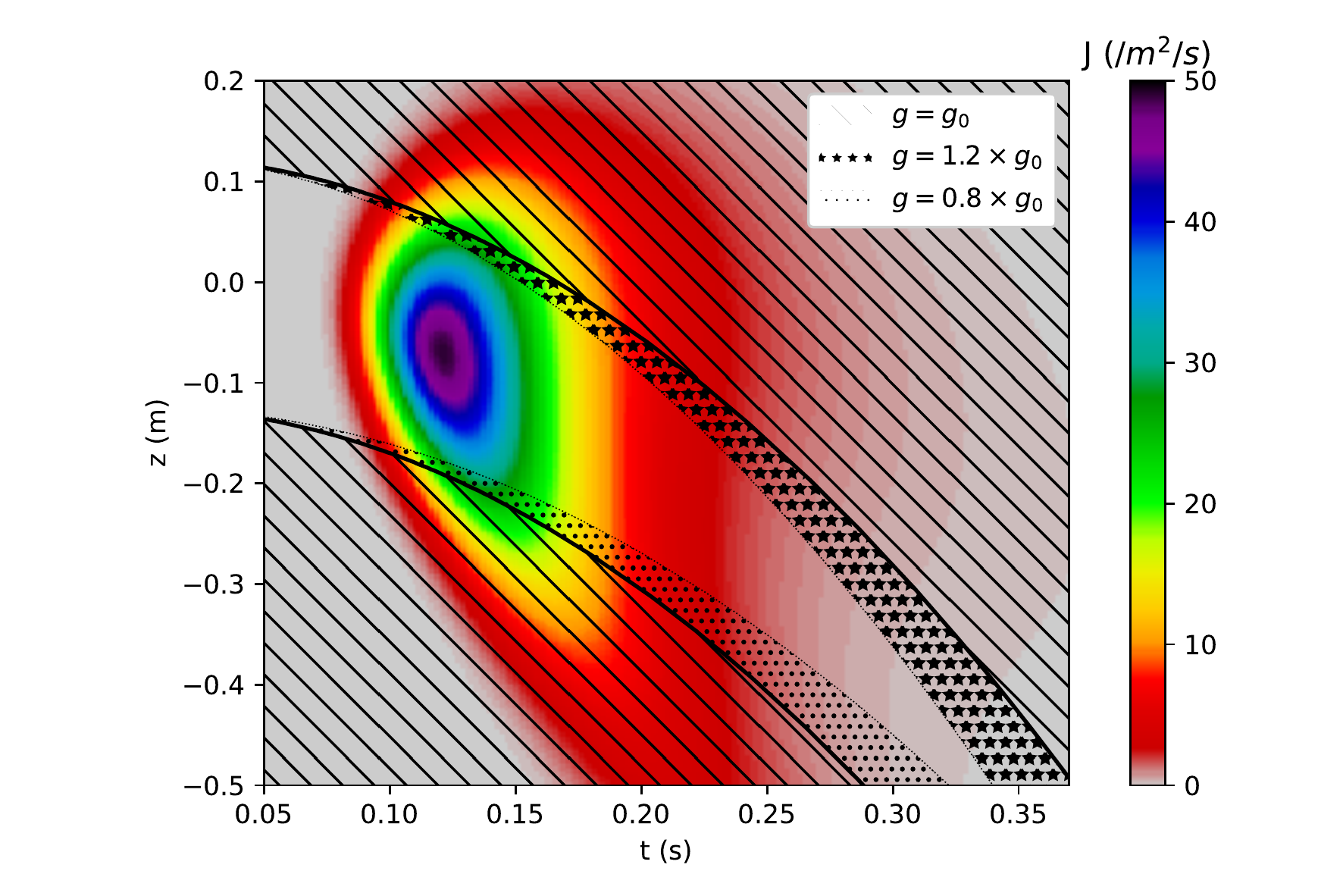}
\caption{\label{figure_J_z_T_limit_shadow} Current $j$ as a function of the altitude $z$ and time of flight $t$. The area with diagonal hatching represents the shadow from the two disks for $g_0$. The two dotted areas represent zones that are not in the shadow for $g=g_0$ but are in the shadow for $g=1.2 \times g_0$ and $g=0.8 \times g_0$. The probability to have an impact in this area correspond the $F_{\max} (1.2 \times g_0)$ and $F_{\min} (0.8 \times g_0)$}
\end{figure}

We now calculate the statistics of $g_{\max}$, the same method being applicable for $g_{\min}$. 
To this aim, we first define the following expectation taken over all possible impact parameters  without obstacles
\begin{eqnarray}
F_{\max} (g) = \mathbb{E}\left(\lambda_{g_0}(1-\lambda_{g})\right)~, \label{eq:F_max}
\end{eqnarray} 
that is the probability to be in the allowed area for $g_0$ and in the shadow for $g$. The function $F_{\max}(g)$, shown on Fig.~\ref{figure_J_z_T_limit_shadow}, is also the cumulative distribution function of $g_{\max}$ for a single event and $f_{\max}(g) = F^\prime_{\max}(g)$, the distribution function. For a draw of $N$ events, the distribution of $g_{\max}$ can then be written as
\begin{equation}
f_{N,\max}(g) = N f_{\max} (g) (1-F_{\max}(g))^{N-1} ~.
\end{equation}

For large value of $N$ and for $(g-g_0)f_{\max} (g_0)\ll 1$, the random variable $(g_{\max}  - g_0)$ follows an exponential distribution of parameters $N f_{\max}(g_0)$
\begin{equation}
f_{N, {\max} }(g) \simeq N f_{\max} (g_0) e^{-N(g-g_0)f_{\max} (g_0)} ~.
\end{equation}
The expected value of $g_{\max}$ is $g_0 + 1/(N f_{\max} (g_0))$ and its standard deviation $1/(N f_{\max} (g_0))$.

The likelihood $\cL$ with obstacles is the product of the likelihood $\cL_0$ without obstacles and the rectangular function $\pi$ limited by $g_{\min}$ and $g_{\max}$. The width of the $\cL_0$ scales as $1/\sqrt{N}$ while the width of $\pi$ scales as 
$1/N\left(1/(f_{\max} (g_0)) + 1/(f_{\min} (g_0))\right)$.
When the width of $\pi$ dominates the final shape, the likelihood function $\cL$ has a trapezoidal shape. By disregarding the slope of the plateau of the trapeze, we can approximate the estimator $\check g$ as 
(the label $\mathrm{mm}$ stands for ``min-max'')
\begin{equation}
\check g_\mathrm{mm} \equiv \frac{g_{\max}  + g_{\min} }2~.
\label{minmaxestimator}
\end{equation}

For large enough values of $N$, the events that contribute to $g_{\max} $ and $g_{\min} $ are different since they correspond to events close to different edges and there is no overlap between the two dotted areas on Fig.~\ref{figure_J_z_T_limit_shadow}. When this is the case, we can assume that $g_{\max} $ and $g_{\min} $ are uncorrelated variables and thus get the following expectation and variance for the min-max estimator \eqref{minmaxestimator}
\begin{eqnarray} 
&&\mathbb E(\check g_\mathrm{mm}) = g_0+\frac 1{2N}\left(\frac 1{f_{\max} (g_0)} - \frac 1{f_{\min} (g_0)}\right) ~, \notag\\
&&\mathbb V\mathrm{ar}(\check g_\mathrm{mm}) = \frac 1{4N^2}\left(\frac 1{f_{\max} (g_0)^2} + \frac 1{f_{\min} (g_0)^2} \right) ~.\label{eq:min_max_model}
\end{eqnarray}

With the two disks symmetrically positioned with respect to the centre of the trap and sufficiently close to it, $f_{\max} (g_0)=f_{\min} (g_0)$ and we call this quantity $f_\mathrm{mm}(g_0)$. The estimator $\check g_\mathrm{mm}$ is then unbiased and its distribution is a Laplace distribution 
\begin{eqnarray}
f_{\check g_\mathrm{mm}}(g) = \frac1{\sqrt{2}\sigma_\mathrm{mm}} e^{-\frac{|g-g_0|}{\sqrt{2}\sigma_\mathrm{mm}}} ~,\notag\\
\sigma_\mathrm{mm} = \frac1{\sqrt 2N f_\mathrm{mm}(g_0)}~,
\end{eqnarray}
where we have denoted $\sigma_\mathrm{mm}$ the dispersion of $\check g_\mathrm{mm}$ which scales as $1/N$.



\subsection{The Cramer-Rao bound}
\label{subsec:thecramerraobound}

We now discuss the cases of large values of $N$ for which the dispersion scales as $1/\sqrt{N}$ and can be approximated by a Cramer-Rao bound \cite{Frechet,Cramer,Refregier}. This corresponds to the limit of an efficient sampling of the edges and imposes to account for the rounding up in $J$ of the steps in $j$, thanks to the dispersion $\tau$ of $t_0$ (see eq.\eqref{eq:logistic}). 

Due to the small value of $\tau$ compared to the time scale in $j$, the convolution does not change appreciably the current, except in the vicinity of the steps. An approximation of $J$ is thus given by the expressions 
\begin{eqnarray}
&&J_\tau(\bR , T) = J_0(\bR , T) \Lambda_{\tau, g}(\bR , T)
~, \label{eq:J_tau_app} \\
&&\Lambda_{\tau, g}(\bR , T) =  \int \lambda_g\left(\bR , T-u\right) \delta _{\tau }\left(u\right) \md u~.
\notag
\end{eqnarray}

Using equation~\eqref{eq:J_tau_app}, one can decompose the integral giving the Fisher information as a sum of 3 terms 
\begin{eqnarray}
&&\cI_g = \cI_{g,1} + \cI_{g,2} + \cI_{g,3} ~, \\
&&\cI_{g,1} = \int \md S \md T~\frac{\left( \partial _{g}J_0\right) ^{2}}{J_0}\Lambda_\tau ~,    \notag\\
&&\cI_{g,2} = \int \md S \md T~\frac{\left( \partial _{g}\Lambda_\tau\right) ^{2}}{\Lambda_\tau}J_0 ~, \notag\\
&&\cI_{g,3} = 2 \int \md S \md T ~\left( \partial _{g}\Lambda_\tau\right)\left( \partial _{g}J_0\right) ~.
\notag
\end{eqnarray}
The first term $\cI_{g,1}$ is the Fisher integral for the current calculated without obstacles. The second term is the Fisher information added by the steps in the current $j$. For this term, the integrand is non negligible only in the vicinity of the step, with the extra information arising because the position of the step depends on $g$. 

We first calculate this integral for a single step at a time $t_s$. In this case, $\lambda(t) = \theta(t - t_s)$ where $\theta$ is the Heaviside function and $\Lambda_\tau(T) = \theta_\tau(T - t_s)$ where $\theta_\tau(T) = \int_{-\infty}^{T} \delta_\tau(u)\md u$. Neglecting the variation of the current $j_0$ over the step, this integral can be obtained analytically 
\begin{equation}
\int \text{d}T\frac{\left( \delta _{\tau }\left( t_\mathrm{s}
-T\right) \right) ^{2}}{\theta _{\tau }\left( t_{\mathrm{s}}-T\right) }= \frac{1}{2\tau }  ~.
\end{equation}
This integral has been calculated exactly for the logistic distribution \eqref{eq:logistic} of photodetachement moment, but similar scaling laws would hold for other models of rounded step, with the integrals being a definition of $\tau $. Finally $\cI_{g,3}$ contains all other terms, it scales as $\tau ^0$ and is thus negligible with respect to $\cI_{g,2}$.

\begin{figure}[t!]
\includegraphics[scale=0.64]{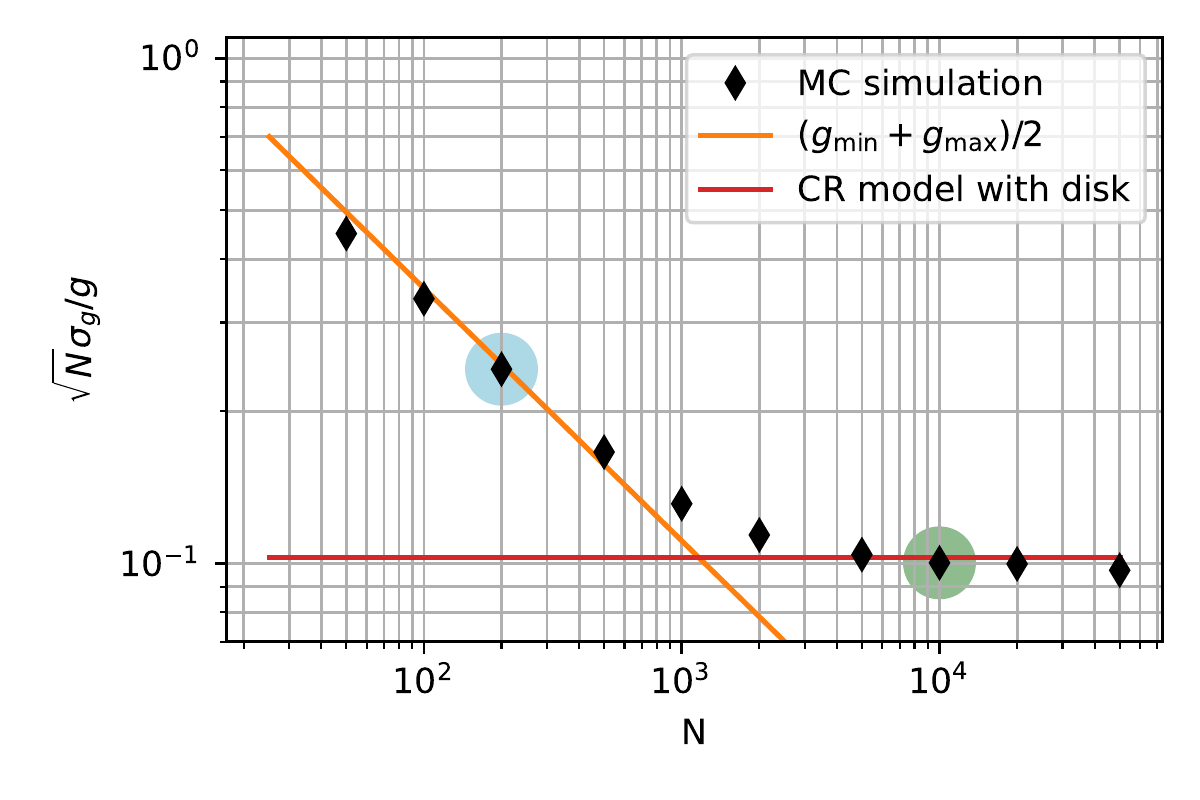}
\caption{Relative dispersions obtained with Monte-Carlo (MC) simulations multiplied by $\sqrt{N}$ and represented versus $N$ (black diamonds). The dark red line represents the Cramer-Rao (CR) bound calculated with obstacles. The orange line is the limit given by the min-max model of eq.\eqref{eq:min_max_model}. Results are drawn for default configuration for the disks $R_\md=\SI{2}{\centi\meter}$, $H_\md=\SI{1}{\centi\meter}$ and $\tau=\SI{200}{\micro\second}$). Highlighted in blue and green are the points whose normalised histograms are shown on Fig.~\ref{fig:histogramvsN}.}
\label{fig:courbe_pierre} 
\end{figure}

We thus obtain a large contribution scaling as $\frac1\tau$
\begin{equation}
\cI_{g,2}\simeq \int \md S~\frac{J_0(t_s)}{2\tau }
~\left( \partial _{g}t_{\mathrm{s}}\right) ^{2}  ~.
\label{eq:fisher_un_sur_tau}
\end{equation}
In order to ease the calculation of the large $\mathcal{I}_{g,2}$, an approximation of $\partial_{g}t_{\mathrm{s}}$ can be made when obstacles are close to the source and gravity can be neglected during the flight between the source and the disks, which leads to the simple relationship
\begin{equation}
\partial_{g}t_{\mathrm{s}} \simeq \frac{t_s}{2g}~.
\end{equation}
This relationship can be seen on Fig.~\ref{figure_J_z_T_limit_shadow} where the horizontal width of the dotted area (proportional to $\partial_g t_s$) is proportional to $t$. 

Equation \eqref{eq:fisher_un_sur_tau} gives an integral on the boundary defined by the function $\lambda_g$. It can be replaced by an integral over the volume of points that are within $\delta g/2$ from the boundary
\begin{equation}
\cI_{g,2} \simeq \frac1{4g_0\tau} \frac{\mathbb E_{g_0}\left(t\left|\lambda_{g_0-\delta g/2} - \lambda_{g_0+\delta g/2}\right|\right)}{|\delta g|}~.
\label{eq:model_un_sur_tau}
\end{equation}
Here the expectation used to represent the integral is taken for impacts without obstacles.
Choosing an adequate value of $\delta g$, this formula can be used to numerically calculate $\cI_{g,2}$ with a Monte-Carlo method when no analytical formula for 
$t_s(\bR )$ is available. 

These discussions show that the large contribution $\cI_{g,2}$ to Fisher information is proportional to the density of events close to the step weighted by the time of flight (the longer it is, the higher is the information). Note also that the expected value is taken on both sides of the step. This formula is similar to \eqref{eq:F_max} used to compute $f_{\max} $ and leads to a rough relationship between the 2 quantities
\begin{equation}
\frac{g\cI_{g,2}}{f_\mathrm{mm}} \approx \frac{\left<T\right>}{2\tau}~,
\label{eq:rough_value_I}
\end{equation}
where $\left<T\right>$ is a typical value of the time of flight ($R_\mc/v$), with $R_\mc$ the radius of the chamber (see Fig.\ref{fig:trap}).

\begin{figure}
\includegraphics[width=.95\linewidth]{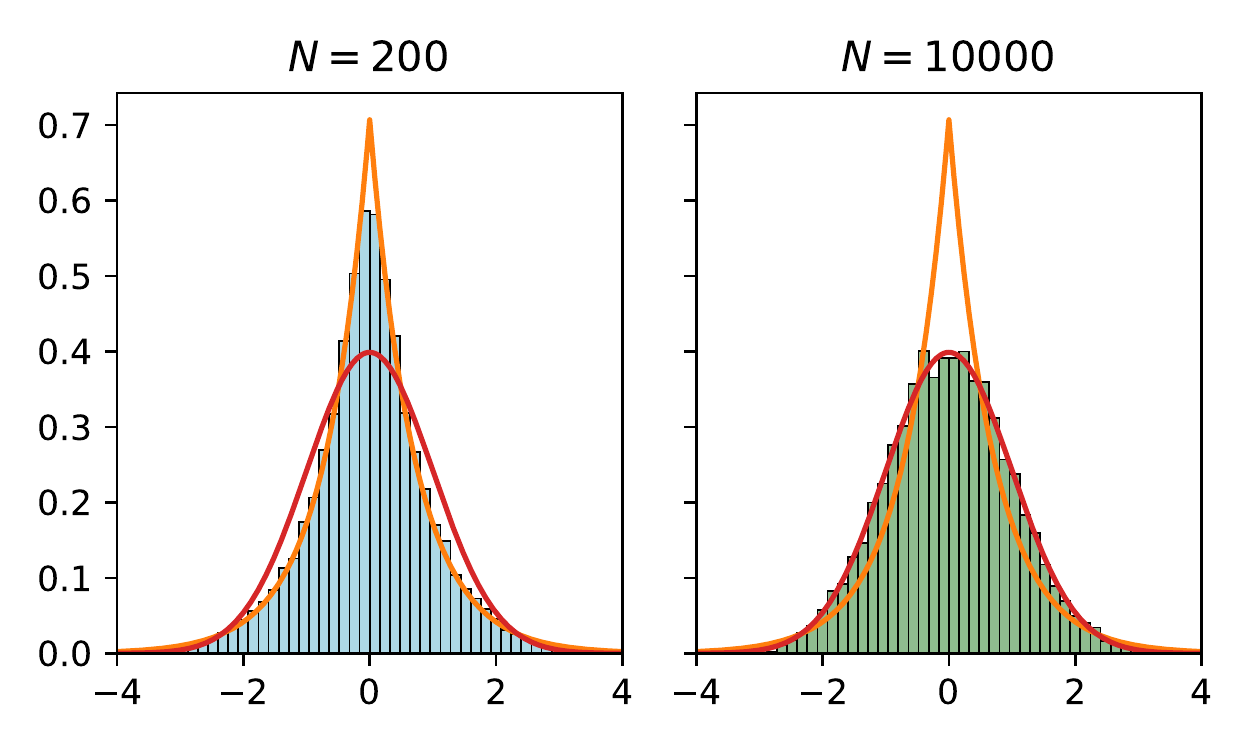}
\caption{Standardised histograms of $\check g$ for the points $N=200$ and $10000$ highlithed (with the same colors) on Fig.~\ref{fig:courbe_pierre}. The number of draws is $50000$ for $N=200$ and $12000$ for $N=10000$. The red curves represent normal distributions, a fair approximation in the limit of efficient statistics, and the orange curve Laplace distributions, a fair approximation in the min-max model. }
\label{fig:histogramvsN}
\end{figure}

When there are more than one step at a position $\bR $, the Fisher integral is the sum over all steps of contributions \eqref{eq:fisher_un_sur_tau} or \eqref{eq:model_un_sur_tau}.

\subsection{Discussion}
\label{subsec:discussion}

In order to assess the qualitative results of the min-max and Cramer-Rao models presented in \S \ref{subsec:theminmaxmodel} and \ref{subsec:thecramerraobound}, we have performed full Monte Carlo simulations of the experiment for different sets of parameters and for a simple geometry without the floor and the ceiling of the chamber so that atoms are detected only on the walls. The results of the simulation are shown on Fig.~\ref{fig:courbe_pierre}. 

Simulations correspond to the default configuration for the disks ($R_\md=\SI{2}{\centi\meter}$, $H_\md=\SI{1}{\centi\meter}$) considered for all other figures and to a starting time dispersion $\tau=\SI{200}{\micro\second}$. Black diamonds represent the relative standard deviation calculated without approximation and multiplied by $\sqrt{N}$ as a function of $N$. 
Two limits are also plotted: the Cramer-Rao bounds with obstacles (red line) and the min-max model where $\check g$ is estimated from eq.\eqref{eq:min_max_model} (orange line).

We note that the min-max model gives an approximation of the dispersion even for small numbers $N$. The transition between the min-max model (where the dispersion is independent of $\tau$) and the Cramer-Rao bound is observed for $N\approx 1000$ with $\tau=\SI{200}{\micro\second}$. Using eq.\eqref{eq:rough_value_I}, the intersection $N_*$ of the two curves is found to be
\begin{equation}
N_* \simeq \frac{\left<T\right>}{2\tau}\frac{1}{g f_\mathrm{mm}}~.
\label{eq:limit_N}
\end{equation}
This equation essentially tells us that the Cramer-Rao bound is reached when the number of atoms within a delay of less than $\tau$ from the step crosses unity.

\begin{figure}
\includegraphics[width=.95\linewidth]{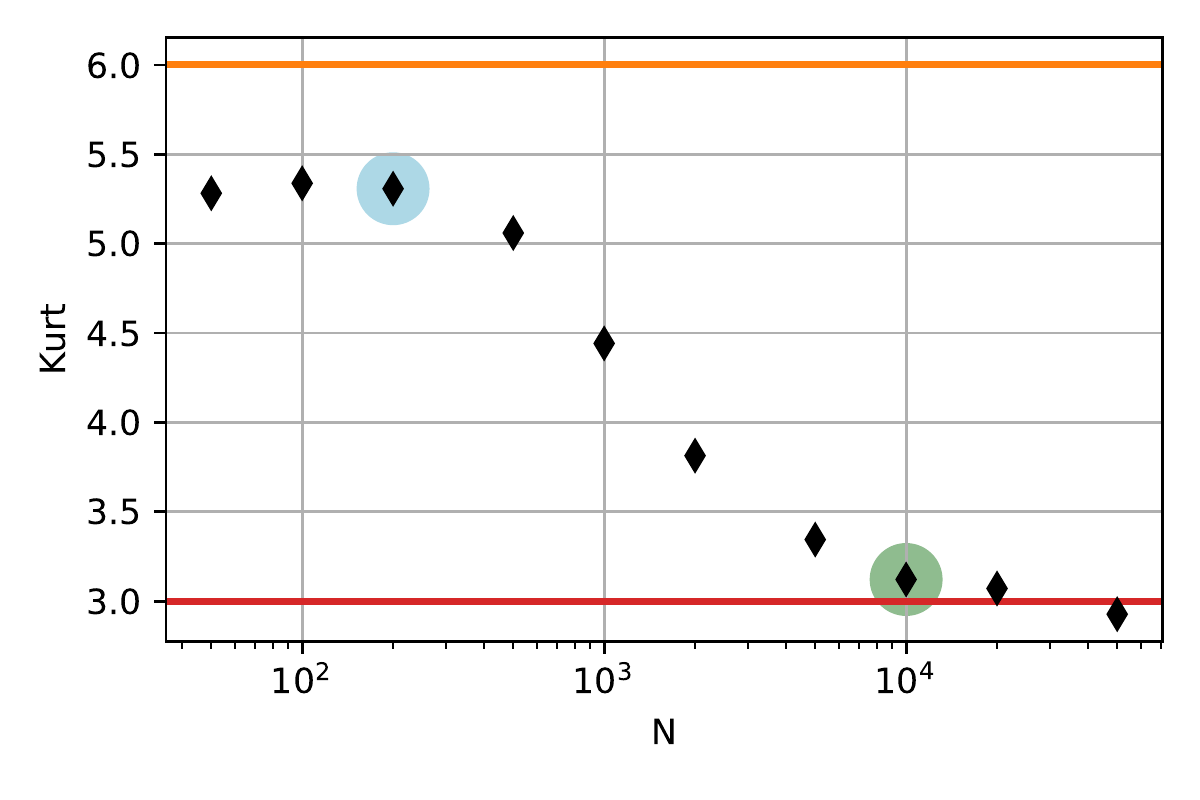}
\caption{Kurtosis of the histograms of $\check g$ in the configuration of Fig.~\ref{fig:courbe_pierre}\textbf{b}, drawn as a function of $N$. The horizontal lines represent the limiting values of 3 and 6 obtained respectively for Gauss and Laplace distributions. Highlighted points correspond to those shown on Figs.~\ref{fig:courbe_pierre}-\ref{fig:histogramvsN}.}
\label{fig:kurtosisvsN}
\end{figure}

An interesting way to assess the quality of the analysis is to look at histograms of estimators shown on Fig.\ref{fig:histogramvsN} for 2 different values of $N$ (same parameters as on Fig.~\ref{fig:courbe_pierre}). When the sampling of the edge is efficient, the distribution of estimators tends to have a Gaussian shape indicated by the red curves on Fig.\ref{fig:histogramvsN}. In the opposite case, the statistics tends to be given by the non-gaussian min-max model, so that the distribution of estimators tends to fit a Laplace distribution indicated by the orange curves on Fig.\ref{fig:histogramvsN}. These predictions of the simple models are approximately met by the results of full simulations, with distributions for $N=20$ and $N=100000$ approaching respectively Gauss and Laplace shapes. 

A quantitative assessment of the shape of the distribution is the kurtosis which should be 3 for a Gauss shape and 6 for a Laplace shape. Fig.\ref{fig:kurtosisvsN} shows the variation of the kurtosis for the parameters corresponding to the black diamonds on Fig.\ref{fig:courbe_pierre}. 
When the sampling of the edge is efficient (\textit{ie} for large values of $N$), the distribution tends to have a Gauss shape and the kurtosis effectively approaching 3 (red line). When the statistics is in contrast dominated by the non-gaussian min-max model, the distribution tends to have a Laplace shape and the kurtosis approaches 6 (orange line). 

\begin{figure}[tb!]
\center\includegraphics[width=.8\linewidth]{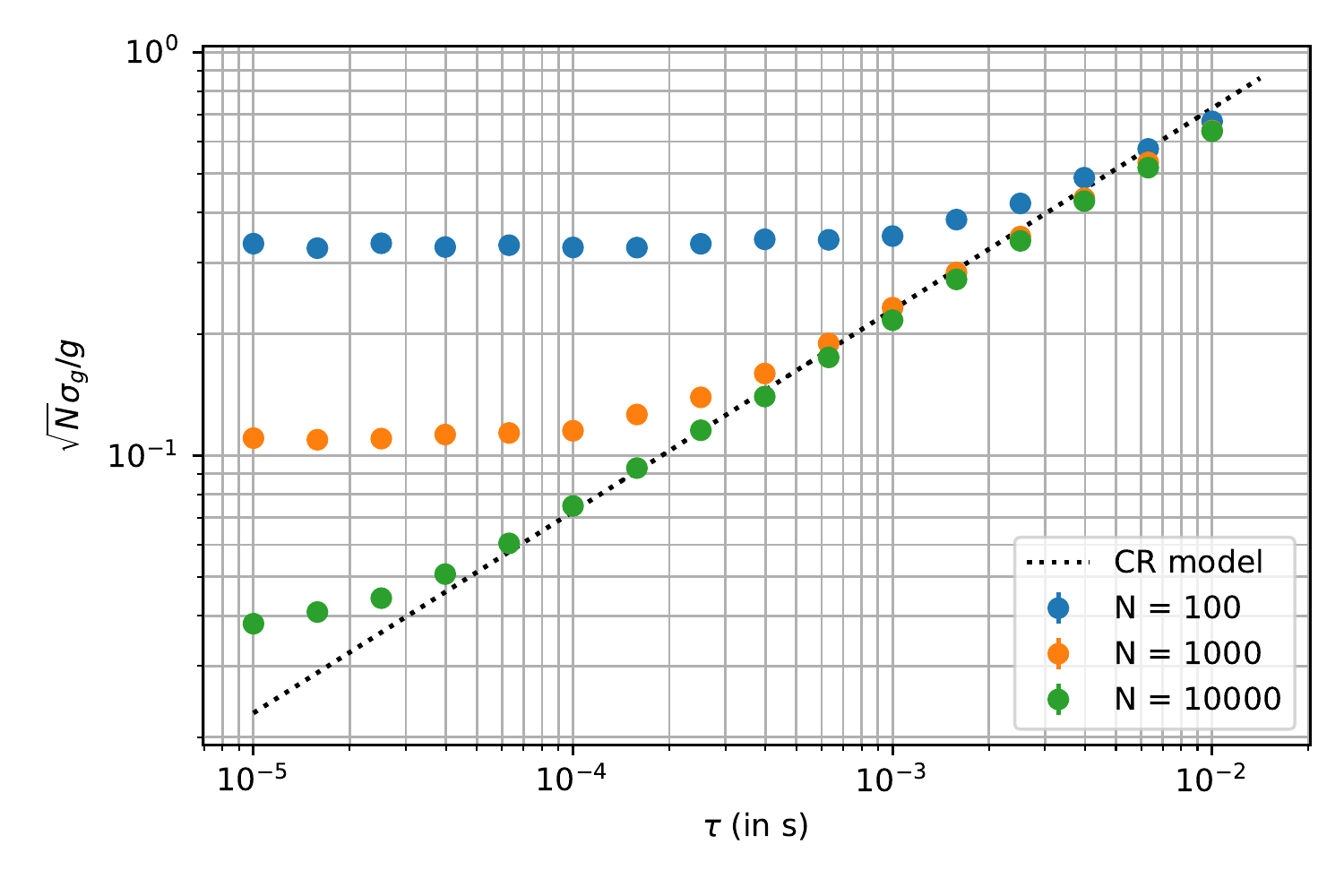}
\caption{Relative dispersion $\sigma$ multiplied by $\sqrt{N}$ as a function of $\tau$ for different values of $N$. For small values of $\tau$, this quantity tends to a finite value depending on $N$. For large values of $\tau$, the Cramer-Rao bound is reached and the quantity tends to a universal curve, meaning that $\sigma$ scales as $\sqrt{\tau/N}$. The dotted line represents the Cramer-Rao limit obtained using the Fisher information given in eq.\eqref{eq:model_un_sur_tau}.}
\label{fig:courbe_pierre_tau}
\end{figure}

We have finally shown on Fig.~\ref{fig:courbe_pierre_tau}  the standard deviation multiplied by $\sqrt{N}$ as a function of the dispersion $\tau$. This quantity tends to a limit independent of $N$ for large values of $\tau$, which corresponds to a statistical efficiency close to 1 and a dispersion approaching the Cramer-Rao bound. On the other hand, the  Fisher information scales as $1/\tau$ for small values of $\tau$, where the estimator is efficient only for very large value of $N$. For $\tau$ around \num{1E-3} s, there is a good agreement with the model of equation \eqref{eq:model_un_sur_tau}.

\section{Effect of quantum reflections}
\label{sec:effect}

Ultra-cold anti-hydrogen atoms falling onto the detection plate suffer a quantum reflection (QR) on the Casimir-Polder potential before touching the surface and this could affect the free fall measurement  \cite{Dufour2013,Dufour2015}. As quantum reflections could change the results of the discussions presented up to now, we have repeated the analysis by taking QR into account.

The probability of quantum reflection on the plate depends on the component of velocity orthogonal to the plate and on the optical properties of the material. 
Here we assume that the boundaries of the free fall chamber are well polished stainless steel plates behaving as mirror of good optical quality. A good approximation of quantum reflection probabilities is thus obtained by taking the values calculated for a mirror perfectly reflecting electromagnetic fields \cite{Dufour2013}.

For simplicity, we use an interpolation formula which has been designed to reproduce accurately the full range of numerically calculated curves (Dufour G., Guérout G., Lambrecht A. and Reynaud S., private communication) 
\begin{eqnarray}
&&|r|^2=\exp\left(-4\kappa \huge{/} \left(1+\frac{\alpha\kappa^{2/3}}{1+\beta\kappa^{-1}}\right)\right)  ~,~  \label{interpolation}
\\
&&\kappa\equiv k|b|=
\frac{m|b|}{\hbar}|V_\perp|~,~
\alpha\simeq 0.7088 ~,~\beta \simeq 0.5163 ~. \notag
\end{eqnarray}
Here $k$ is the atomic wavevector determined by the orthogonal velocity $V_\perp$, and $b$ the imaginary part of the scattering length deduced from the optical properties of the surfaces.
For a mirror perfectly reflecting electromagnetic fields, $|b| \simeq 28.75$nm  \cite{Dufour2014}.
The constants $\alpha$ and $\beta$ have been obtained by a least-squares fit on the numerically calculated curves.
The formula \eqref{interpolation} reproduces the analytical asymptotic behaviours known at low and high energies and it gives an estimate of the reflection probability with a relative dispersion better than $1\times10^{-4}$ at all energies.

\begin{figure}[t!]
\includegraphics[width=0.8\linewidth]{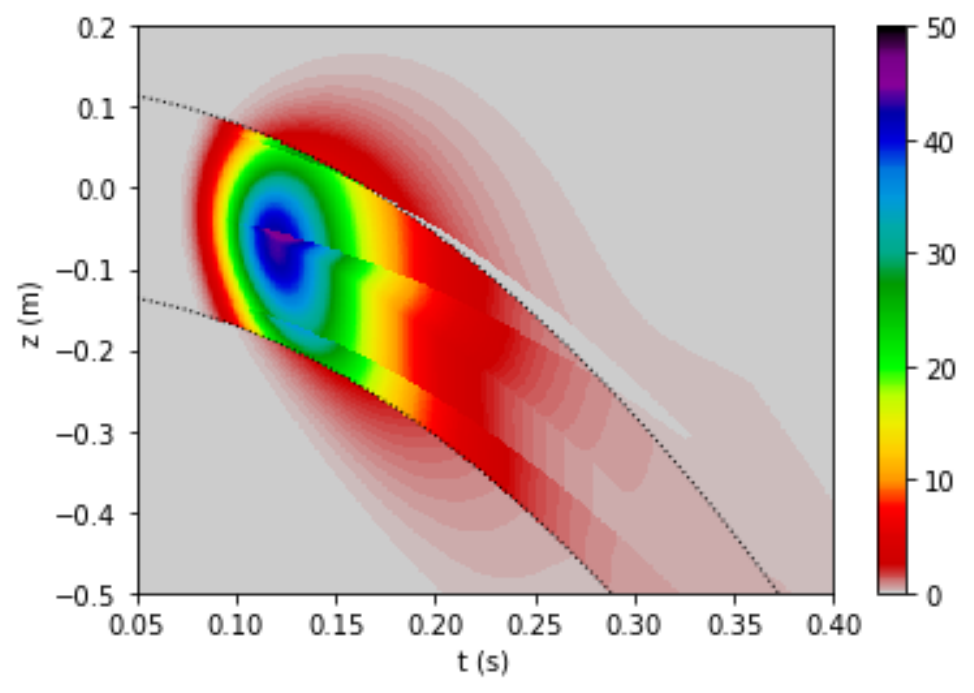} 
\caption{Particle current $J$ incident on the walls calculated taking into account quantum reflections. The limit of the shadow zone is represented in dotted lines.}
\label{fig:current-with-QR}
\end{figure}

As different velocities correspond to neatly different probabilities, it is necessary to calculate quantum reflection probability for each individual trajectory. 
Though quantum reflection probabilities are small, they can give rise to systematics of the same order of magnitude as the statistical accuracy looked for in GBAR experiment, and it is necessary to take them into account in the analysis.

From a detection at positions $(\bR ,T)$ in space and time, we have to find the initial velocity on the trajectory. There is a one to one matching between those values, which however depends on reflections in the interval between initial launch and detection on a surface of the free fall chamber.  
Precisely, a detection point on a surface of the chamber can be reached directly or by having undergone one or several reflections on the disks or on another surface of the chamber. As elementary quantum reflection probabilities are small, we disregard here the case of multiple quantum reflections.

The probability current is obtained by adding the different contributions
\begin{equation}
    J(\bR ,T)=J_{\mathrm{dir}}(\bR ,T)
    +\sum_\mathrm{surf} J_\mathrm{QR}^{(s)}(\bR ,T) ~,
\end{equation}
where $J_{\mathrm{dir}}(\bR ,t)$ corresponds to direct trajectories while each $J_\mathrm{QR}^{(s)}$ describes the case with one quantum reflection on the surface $s$. Each of the latter expressions contains the associated quantum reflection probability.

For the configuration of cylindrical chamber with disks
(same parameters than in the default configuration considered above), with parameters $f=\SI{1}{\mega\hertz}$, $\delta E=\SI{30}{\micro\electronvolt}$ and horizontal polarisation of the laser, the fraction of atoms that reach the surfaces of the detection chamber is about 66\% (the other 34\% are annihilated on the disks and useless for the measurement of $g$) while $\sim$18\% of the atoms annihilated on the surfaces of the free fall chamber have been reflected on another surface before their detection.

We have represented on Fig.\ref{fig:current-with-QR}  the current on the walls as function of time $t$ and position coordinate $z$, with a choice of parameters corresponding to that in Fig.\ref{fig:figure_J_z_T_two_diska} except for the fact that quantum reflection is now accounted for. The essential information on the new plots is that quantum reflections allow atoms to reach the shadow zone which was previously forbidden. We also observe that there remains a small forbidden zone which cannot be reached by any trajectory even when taking into account quantum reflections.

\begin{figure}[t!]
\center\includegraphics[width=.8\linewidth]{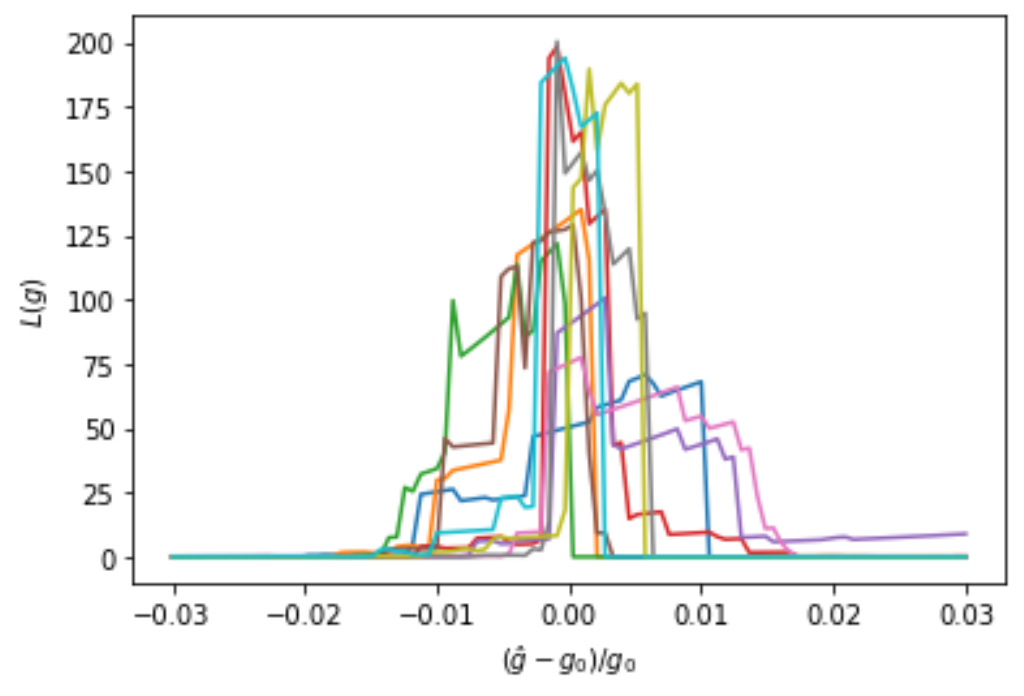}
\caption{Sample of normalised likelihoods including quantum reflections calculated for independent random draws of 1000 atoms. Same parameters as in Fig.\ref{fig:current-with-QR}.}
\label{fig:likelihoods}
\end{figure}

We repeat all steps in calculations described in section \ref{sec:dispersion} now taking into account quantum reflections in the simulation as well as in the estimation stages. 

\begin{figure}[b!]
\center\includegraphics[width=.8\linewidth]{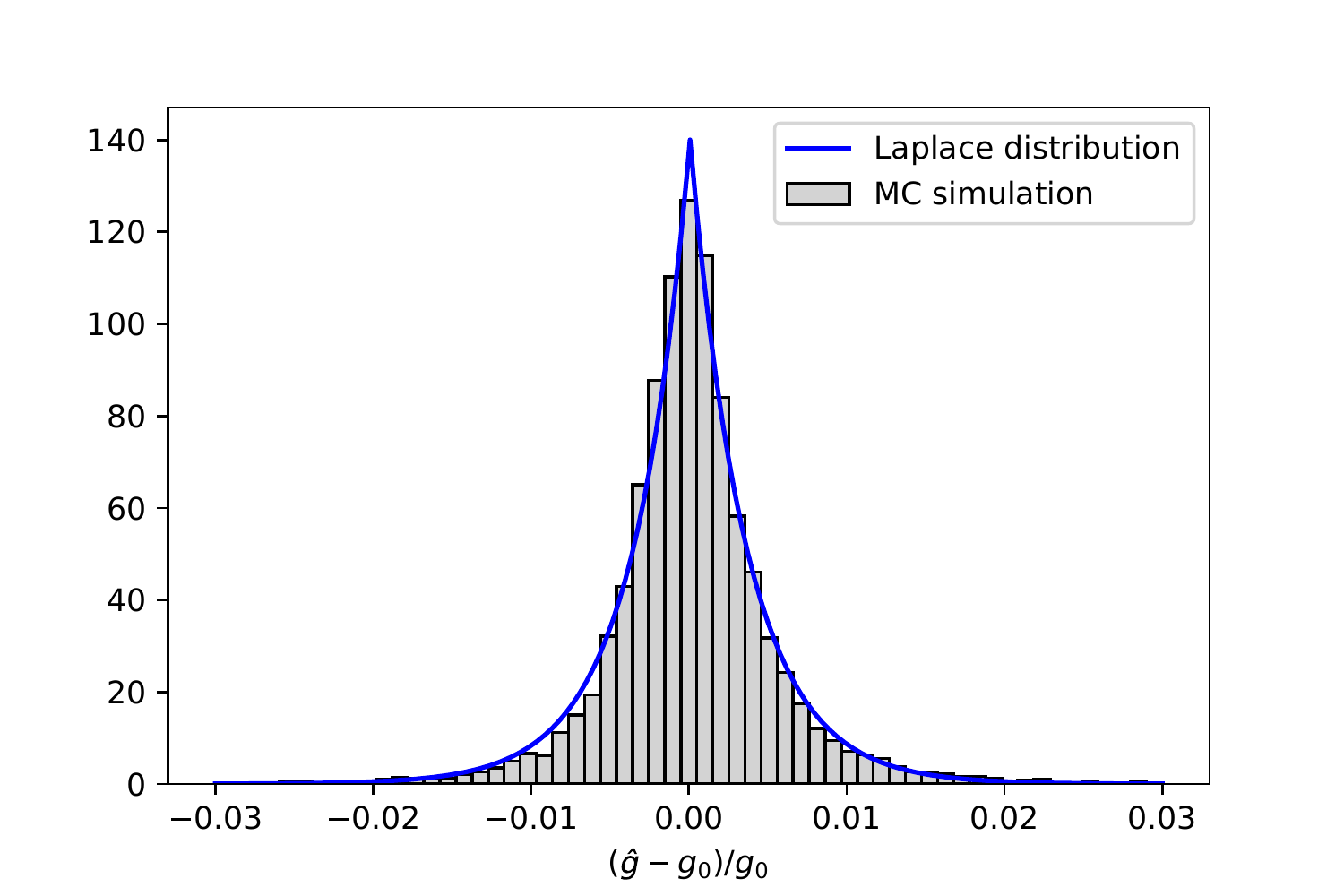}
\caption{Normalised histogram of 10000 estimators $\check g$ obtained with quantum reflection. Same parameters as in Fig.\ref{fig:current-with-QR}.}
\label{fig:histograms_QR}
\end{figure}

We represent on Fig.\ref{fig:likelihoods} the likelihood functions calculated for random draws of $N=1000$ $\Hbar$ atoms. We clearly see that the likelihoods are not Gaussian and contain different steps, in particular due to interception of some trajectories by the disks. We also notice that some likelihood functions are significantly biased.

We then show on Fig.\ref{fig:histograms_QR} an histogram of the estimator $\check g$ obtained by repeating the process presented in section \ref{sec:dispersion}. We deduce the average $\mu_g$ and the standard deviation $\sigma_g$ of the estimators of $g$. The relative statistical bias ($\mu_g-g_0)/g_0$ and relative dispersion ($\sigma_g/g_0$) are found to be respectively $~0.03\%$ and $0.54\%$.

As could be expected, the presence of quantum reflection degrades the dispersion but the degradation is limited when considering that the expected relative dispersion was $0.36\%$ with the same experimental conditions (initial velocity distribution and parameters of the photodetachment laser), with quantum reflection not accounted for. 

For completeness, we also evaluated the confidence intervals containing 95\% of the probability in the histogram of the estimators $\check g$. We found $[9.751 ; 9.868]$ for the confidence interval with no quantum reflection and $[9.739 ; 9.891]$ for the confidence interval  including quantum reflection. As could be expected, the confidence intervals are larger than if they were calculated for a Gaussian distribution with the known standard deviations. However, there is no significant difference in this respect associated with quantum reflection.

\section{Conclusion}
\label{sec:conclusion}

In this paper we have studied in a detailed manner the effect of the obstacles present in the vicinity of the source on the dispersion of the free fall acceleration measurement of $\Hbar$ atoms to be performed by the GBAR experiment. In order to ease the discussion, we have considered a clean geometry with two disks symmetrically positioned above and below the source to hide the obstacles.

We have first performed Monte-Carlo simulations in order to discussed the accuracy to be expected for the measurement. We have shown that the accuracy is improved thanks to the additional information on the value of $g$ gained from the presence of shadow edges the positions of which depend on $g$. We have also developed new studies of the statistics of events close to an edge to obtain a quantitative understanding of the different regimes observed for the variation of the dispersion versus the number $N$ of 
$\Hbar$ atoms. 

We have finally taken into account quantum reflection processes on the Casimir-Polder potential above matter surfaces. These processes lead to detection of $\Hbar$ atoms in the shadow zones, which could have been detrimental for the accuracy. We have however shown that quantum reflection only slightly reduces the advantage coming from the gain of information associated with shadow edges.

\paragraph*{Acknowledgements}
We thank our colleagues in the GBAR collaboration \cite{GBARpage} for insightful discussions, in particular F. Biraben, P.P. Blumer, P. Crivelli, P. Debu, A. Douillet, N. Garroum, L. Hilico, P. Indelicato, G. Janka, J.-P. Karr, L. Liszkay, B. Mansouli\'e, V.V. Nesvizhevsky, F. Nez, N. Paul, P. P\'erez, C. Regenfus, F. Schmidt-Kaler, A.Yu. Voronin, S. Wolf.
This work was supported by the Programme National GRAM of CNRS/INSU with INP and IN2P3 co-funded by CNES.

\bibliography{gbar}

\providecommand{\noopsort}[1]{}\providecommand{\singleletter}[1]{#1}%
\begin{thebibliography}{27}%
\makeatletter
\providecommand \@ifxundefined [1]{%
 \@ifx{#1\undefined}
}%
\providecommand \@ifnum [1]{%
 \ifnum #1\expandafter \@firstoftwo
 \else \expandafter \@secondoftwo
 \fi
}%
\providecommand \@ifx [1]{%
 \ifx #1\expandafter \@firstoftwo
 \else \expandafter \@secondoftwo
 \fi
}%
\providecommand \natexlab [1]{#1}%
\providecommand \enquote  [1]{``#1''}%
\providecommand \bibnamefont  [1]{#1}%
\providecommand \bibfnamefont [1]{#1}%
\providecommand \citenamefont [1]{#1}%
\providecommand \href@noop [0]{\@secondoftwo}%
\providecommand \href [0]{\begingroup \@sanitize@url \@href}%
\providecommand \@href[1]{\@@startlink{#1}\@@href}%
\providecommand \@@href[1]{\endgroup#1\@@endlink}%
\providecommand \@sanitize@url [0]{\catcode `\\12\catcode `\$12\catcode
  `\&12\catcode `\#12\catcode `\^12\catcode `\_12\catcode `\%12\relax}%
\providecommand \@@startlink[1]{}%
\providecommand \@@endlink[0]{}%
\providecommand \url  [0]{\begingroup\@sanitize@url \@url }%
\providecommand \@url [1]{\endgroup\@href {#1}{\urlprefix }}%
\providecommand \urlprefix  [0]{URL }%
\providecommand \Eprint [0]{\href }%
\providecommand \doibase [0]{https://doi.org/}%
\providecommand \selectlanguage [0]{\@gobble}%
\providecommand \bibinfo  [0]{\@secondoftwo}%
\providecommand \bibfield  [0]{\@secondoftwo}%
\providecommand \translation [1]{[#1]}%
\providecommand \BibitemOpen [0]{}%
\providecommand \bibitemStop [0]{}%
\providecommand \bibitemNoStop [0]{.\EOS\space}%
\providecommand \EOS [0]{\spacefactor3000\relax}%
\providecommand \BibitemShut  [1]{\csname bibitem#1\endcsname}%
\let\auto@bib@innerbib\@empty
\bibitem [{\citenamefont {Hori}\ and\ \citenamefont {Walz}(2013)}]{Hori2013}%
  \BibitemOpen
  \bibfield  {author} {\bibinfo {author} {\bibfnamefont {M.}~\bibnamefont
  {Hori}}\ and\ \bibinfo {author} {\bibfnamefont {J.}~\bibnamefont {Walz}},\
  }\bibfield  {title} {\bibinfo {title} {{Physics at CERN's Antiproton
  Decelerator}},\ }\href
  {https://doi.org/https://doi.org/10.1016/j.ppnp.2013.02.004} {\bibfield
  {journal} {\bibinfo  {journal} {Progress in Particle and Nuclear Physics}\
  }\textbf {\bibinfo {volume} {72}},\ \bibinfo {pages} {206} (\bibinfo {year}
  {2013})}\BibitemShut {NoStop}%
\bibitem [{\citenamefont {Bertsche}\ \emph {et~al.}(2015)\citenamefont
  {Bertsche}, \citenamefont {Butler}, \citenamefont {Charlton},\ and\
  \citenamefont {Madsen}}]{Bertsche2015}%
  \BibitemOpen
  \bibfield  {author} {\bibinfo {author} {\bibfnamefont {W.~A.}\ \bibnamefont
  {Bertsche}}, \bibinfo {author} {\bibfnamefont {E.}~\bibnamefont {Butler}},
  \bibinfo {author} {\bibfnamefont {M.}~\bibnamefont {Charlton}},\ and\
  \bibinfo {author} {\bibfnamefont {N.}~\bibnamefont {Madsen}},\ }\bibfield
  {title} {\bibinfo {title} {Physics with antihydrogen},\ }\href
  {https://doi.org/10.1088/0953-4075/48/23/232001} {\bibfield  {journal}
  {\bibinfo  {journal} {Journal of Physics B: Atomic, Molecular and Optical
  Physics}\ }\textbf {\bibinfo {volume} {48}},\ \bibinfo {pages} {232001}
  (\bibinfo {year} {2015})}\BibitemShut {NoStop}%
\bibitem [{\citenamefont {Charlton}\ \emph {et~al.}(2017)\citenamefont
  {Charlton}, \citenamefont {Mills},\ and\ \citenamefont
  {Yamazaki}}]{Charlton2017}%
  \BibitemOpen
  \bibfield  {author} {\bibinfo {author} {\bibfnamefont {M.}~\bibnamefont
  {Charlton}}, \bibinfo {author} {\bibfnamefont {A.~P.}\ \bibnamefont
  {Mills}},\ and\ \bibinfo {author} {\bibfnamefont {Y.}~\bibnamefont
  {Yamazaki}},\ }\bibfield  {title} {\bibinfo {title} {Special issue on
  antihydrogen and positronium},\ }\href
  {https://doi.org/10.1088/1361-6455/aa75d8} {\bibfield  {journal} {\bibinfo
  {journal} {Journal of Physics B: Atomic, Molecular and Optical Physics}\
  }\textbf {\bibinfo {volume} {50}},\ \bibinfo {pages} {140201} (\bibinfo
  {year} {2017})}\BibitemShut {NoStop}%
\bibitem [{\citenamefont {Yamazaki}(2020)}]{Yamazaki2020}%
  \BibitemOpen
  \bibfield  {author} {\bibinfo {author} {\bibfnamefont {Y.}~\bibnamefont
  {Yamazaki}},\ }\bibfield  {title} {\bibinfo {title} {Cold and stable
  antimatter for fundamental physics},\ }\href
  {https://doi.org/10.2183/pjab.96.034} {\bibfield  {journal} {\bibinfo
  {journal} {Proceedings of the Japan Academy, Series B}\ }\textbf {\bibinfo
  {volume} {96}},\ \bibinfo {pages} {471} (\bibinfo {year} {2020})}\BibitemShut
  {NoStop}%
\bibitem [{\citenamefont {Collaboration}(2013)}]{Alpha2013}%
  \BibitemOpen
  \bibfield  {author} {\bibinfo {author} {\bibfnamefont {A.}~\bibnamefont
  {Collaboration}},\ }\bibfield  {title} {\bibinfo {title} {Description and
  first application of a new technique to measure the gravitational mass of
  antihydrogen},\ }\href {https://doi.org/10.1038/ncomms2787} {\bibfield
  {journal} {\bibinfo  {journal} {Nature Communications}\ }\textbf {\bibinfo
  {volume} {4}},\ \bibinfo {eid} {1785} (\bibinfo {year} {2013})}\BibitemShut
  {NoStop}%
\bibitem [{\citenamefont {Maury}\ \emph {et~al.}(2014)\citenamefont {Maury},
  \citenamefont {Oelert}, \citenamefont {Bartmann}, \citenamefont
  {Belochitskii}, \citenamefont {Breuker}, \citenamefont {Butin}, \citenamefont
  {Carli}, \citenamefont {Eriksson}, \citenamefont {Pasinelli},\ and\
  \citenamefont {Tranquille}}]{Maury2014}%
  \BibitemOpen
  \bibfield  {author} {\bibinfo {author} {\bibfnamefont {S.}~\bibnamefont
  {Maury}}, \bibinfo {author} {\bibfnamefont {W.}~\bibnamefont {Oelert}},
  \bibinfo {author} {\bibfnamefont {W.}~\bibnamefont {Bartmann}}, \bibinfo
  {author} {\bibfnamefont {P.}~\bibnamefont {Belochitskii}}, \bibinfo {author}
  {\bibfnamefont {H.}~\bibnamefont {Breuker}}, \bibinfo {author} {\bibfnamefont
  {F.}~\bibnamefont {Butin}}, \bibinfo {author} {\bibfnamefont
  {C.}~\bibnamefont {Carli}}, \bibinfo {author} {\bibfnamefont
  {T.}~\bibnamefont {Eriksson}}, \bibinfo {author} {\bibfnamefont
  {S.}~\bibnamefont {Pasinelli}},\ and\ \bibinfo {author} {\bibfnamefont
  {G.}~\bibnamefont {Tranquille}},\ }\bibfield  {title} {\bibinfo {title}
  {{ELENA: the extra low energy anti-proton facility at CERN}},\ }\href
  {https://doi.org/10.1007/s10751-014-1067-y} {\bibfield  {journal} {\bibinfo
  {journal} {Hyperfine Interactions}\ }\textbf {\bibinfo {volume} {229}},\
  \bibinfo {pages} {105} (\bibinfo {year} {2014})}\BibitemShut {NoStop}%
\bibitem [{\citenamefont {Bertsche}(2018)}]{Bertsche2018}%
  \BibitemOpen
  \bibfield  {author} {\bibinfo {author} {\bibfnamefont {W.~A.}\ \bibnamefont
  {Bertsche}},\ }\bibfield  {title} {\bibinfo {title} {Prospects for comparison
  of matter and antimatter gravitation with {ALPHA-g}},\ }\href
  {https://doi.org/10.1098/rsta.2017.0265} {\bibfield  {journal} {\bibinfo
  {journal} {Philosophical Transactions of the Royal Society A: Mathematical,
  Physical and Engineering Sciences}\ }\textbf {\bibinfo {volume} {376}},\
  \bibinfo {pages} {20170265} (\bibinfo {year} {2018})}\BibitemShut {NoStop}%
\bibitem [{\citenamefont {Pagano}\ and\ \citenamefont
  {al.}(2020)}]{Pagano2020}%
  \BibitemOpen
  \bibfield  {author} {\bibinfo {author} {\bibfnamefont {D.}~\bibnamefont
  {Pagano}}\ and\ \bibinfo {author} {\bibnamefont {al.}},\ }\bibfield  {title}
  {\bibinfo {title} {Gravity and antimatter: the {AEgIS} experiment at
  {CERN}},\ }\href {https://doi.org/10.1088/1742-6596/1342/1/012016} {\bibfield
   {journal} {\bibinfo  {journal} {Journal of Physics: Conference Series}\
  }\textbf {\bibinfo {volume} {1342}},\ \bibinfo {pages} {012016} (\bibinfo
  {year} {2020})}\BibitemShut {NoStop}%
\bibitem [{\citenamefont {Mansouli{\'e}}\ and\ \citenamefont {{on behalf of the
  GBAR Collaboration}}(2019)}]{Mansoulie2019}%
  \BibitemOpen
  \bibfield  {author} {\bibinfo {author} {\bibfnamefont {B.}~\bibnamefont
  {Mansouli{\'e}}}\ and\ \bibinfo {author} {\bibnamefont {{on behalf of the
  GBAR Collaboration}}},\ }\bibfield  {title} {\bibinfo {title} {Status of the
  gbar experiment at cern},\ }\href {https://doi.org/10.1007/s10751-018-1550-y}
  {\bibfield  {journal} {\bibinfo  {journal} {Hyperfine Interactions}\ }\textbf
  {\bibinfo {volume} {240}},\ \bibinfo {pages} {11} (\bibinfo {year}
  {2019})}\BibitemShut {NoStop}%
\bibitem [{\citenamefont {Indelicato}\ \emph {et~al.}(2014)\citenamefont
  {Indelicato}, \citenamefont {Chardin}, \citenamefont {Grandemange},
  \citenamefont {Lunney}, \citenamefont {Manea}, \citenamefont {Badertscher},
  \citenamefont {Crivelli}, \citenamefont {Curioni}, \citenamefont
  {Marchionni}, \citenamefont {Rossi}, \citenamefont {Rubbia}, \citenamefont
  {Nesvizhevsky}, \citenamefont {{Brook-Roberge}}, \citenamefont {Comini},
  \citenamefont {Debu}, \citenamefont {Dupr{\'e}}, \citenamefont {Liszkay},
  \citenamefont {Mansouli{\'e}}, \citenamefont {P{\'e}rez}, \citenamefont
  {Rey}, \citenamefont {Reymond}, \citenamefont {Ruiz}, \citenamefont
  {Sacquin}, \citenamefont {Vallage}, \citenamefont {Biraben}, \citenamefont
  {Clad{\'e}}, \citenamefont {Douillet}, \citenamefont {Dufour}, \citenamefont
  {Guellati}, \citenamefont {Hilico}, \citenamefont {Lambrecht}, \citenamefont
  {Gu{\'e}rout}, \citenamefont {Karr}, \citenamefont {Nez}, \citenamefont
  {Reynaud}, \citenamefont {Szabo}, \citenamefont {Tran}, \citenamefont
  {Trapateau}, \citenamefont {Mohri}, \citenamefont {Yamazaki}, \citenamefont
  {Charlton}, \citenamefont {Eriksson}, \citenamefont {Madsen}, \citenamefont
  {Werf}, \citenamefont {Kuroda}, \citenamefont {Torii}, \citenamefont
  {Nagashima}, \citenamefont {{Schmidt-Kaler}}, \citenamefont {Walz},
  \citenamefont {Wolf}, \citenamefont {Hervieux}, \citenamefont {Manfredi},
  \citenamefont {Voronin}, \citenamefont {Froelich}, \citenamefont {Wronka},\
  and\ \citenamefont {Staszczak}}]{Indelicato2014}%
  \BibitemOpen
  \bibfield  {author} {\bibinfo {author} {\bibfnamefont {P.}~\bibnamefont
  {Indelicato}}, \bibinfo {author} {\bibfnamefont {G.}~\bibnamefont {Chardin}},
  \bibinfo {author} {\bibfnamefont {P.}~\bibnamefont {Grandemange}}, \bibinfo
  {author} {\bibfnamefont {D.}~\bibnamefont {Lunney}}, \bibinfo {author}
  {\bibfnamefont {V.}~\bibnamefont {Manea}}, \bibinfo {author} {\bibfnamefont
  {A.}~\bibnamefont {Badertscher}}, \bibinfo {author} {\bibfnamefont
  {P.}~\bibnamefont {Crivelli}}, \bibinfo {author} {\bibfnamefont
  {A.}~\bibnamefont {Curioni}}, \bibinfo {author} {\bibfnamefont
  {A.}~\bibnamefont {Marchionni}}, \bibinfo {author} {\bibfnamefont
  {B.}~\bibnamefont {Rossi}}, \bibinfo {author} {\bibfnamefont
  {A.}~\bibnamefont {Rubbia}}, \bibinfo {author} {\bibfnamefont
  {V.}~\bibnamefont {Nesvizhevsky}}, \bibinfo {author} {\bibfnamefont
  {D.}~\bibnamefont {{Brook-Roberge}}}, \bibinfo {author} {\bibfnamefont
  {P.}~\bibnamefont {Comini}}, \bibinfo {author} {\bibfnamefont
  {P.}~\bibnamefont {Debu}}, \bibinfo {author} {\bibfnamefont {P.}~\bibnamefont
  {Dupr{\'e}}}, \bibinfo {author} {\bibfnamefont {L.}~\bibnamefont {Liszkay}},
  \bibinfo {author} {\bibfnamefont {B.}~\bibnamefont {Mansouli{\'e}}}, \bibinfo
  {author} {\bibfnamefont {P.}~\bibnamefont {P{\'e}rez}}, \bibinfo {author}
  {\bibfnamefont {J.-M.}\ \bibnamefont {Rey}}, \bibinfo {author} {\bibfnamefont
  {B.}~\bibnamefont {Reymond}}, \bibinfo {author} {\bibfnamefont
  {N.}~\bibnamefont {Ruiz}}, \bibinfo {author} {\bibfnamefont {Y.}~\bibnamefont
  {Sacquin}}, \bibinfo {author} {\bibfnamefont {B.}~\bibnamefont {Vallage}},
  \bibinfo {author} {\bibfnamefont {F.}~\bibnamefont {Biraben}}, \bibinfo
  {author} {\bibfnamefont {P.}~\bibnamefont {Clad{\'e}}}, \bibinfo {author}
  {\bibfnamefont {A.}~\bibnamefont {Douillet}}, \bibinfo {author}
  {\bibfnamefont {G.}~\bibnamefont {Dufour}}, \bibinfo {author} {\bibfnamefont
  {S.}~\bibnamefont {Guellati}}, \bibinfo {author} {\bibfnamefont
  {L.}~\bibnamefont {Hilico}}, \bibinfo {author} {\bibfnamefont
  {A.}~\bibnamefont {Lambrecht}}, \bibinfo {author} {\bibfnamefont
  {R.}~\bibnamefont {Gu{\'e}rout}}, \bibinfo {author} {\bibfnamefont {J.-P.}\
  \bibnamefont {Karr}}, \bibinfo {author} {\bibfnamefont {F.}~\bibnamefont
  {Nez}}, \bibinfo {author} {\bibfnamefont {S.}~\bibnamefont {Reynaud}},
  \bibinfo {author} {\bibfnamefont {I.~C.}\ \bibnamefont {Szabo}}, \bibinfo
  {author} {\bibfnamefont {V.-Q.}\ \bibnamefont {Tran}}, \bibinfo {author}
  {\bibfnamefont {J.}~\bibnamefont {Trapateau}}, \bibinfo {author}
  {\bibfnamefont {A.}~\bibnamefont {Mohri}}, \bibinfo {author} {\bibfnamefont
  {Y.}~\bibnamefont {Yamazaki}}, \bibinfo {author} {\bibfnamefont
  {M.}~\bibnamefont {Charlton}}, \bibinfo {author} {\bibfnamefont
  {S.}~\bibnamefont {Eriksson}}, \bibinfo {author} {\bibfnamefont
  {N.}~\bibnamefont {Madsen}}, \bibinfo {author} {\bibfnamefont
  {D.}~\bibnamefont {Werf}}, \bibinfo {author} {\bibfnamefont {N.}~\bibnamefont
  {Kuroda}}, \bibinfo {author} {\bibfnamefont {H.}~\bibnamefont {Torii}},
  \bibinfo {author} {\bibfnamefont {Y.}~\bibnamefont {Nagashima}}, \bibinfo
  {author} {\bibfnamefont {F.}~\bibnamefont {{Schmidt-Kaler}}}, \bibinfo
  {author} {\bibfnamefont {J.}~\bibnamefont {Walz}}, \bibinfo {author}
  {\bibfnamefont {S.}~\bibnamefont {Wolf}}, \bibinfo {author} {\bibfnamefont
  {P.-A.}\ \bibnamefont {Hervieux}}, \bibinfo {author} {\bibfnamefont
  {G.}~\bibnamefont {Manfredi}}, \bibinfo {author} {\bibfnamefont
  {A.}~\bibnamefont {Voronin}}, \bibinfo {author} {\bibfnamefont
  {P.}~\bibnamefont {Froelich}}, \bibinfo {author} {\bibfnamefont
  {S.}~\bibnamefont {Wronka}},\ and\ \bibinfo {author} {\bibfnamefont
  {M.}~\bibnamefont {Staszczak}},\ }\bibfield  {title} {\bibinfo {title} {The
  {{Gbar}} project, or how does antimatter fall?},\ }\href
  {https://doi.org/10.1007/s10751-014-1019-6} {\bibfield  {journal} {\bibinfo
  {journal} {Hyperfine Interactions}\ }\textbf {\bibinfo {volume} {228}},\
  \bibinfo {pages} {141} (\bibinfo {year} {2014})}\BibitemShut {NoStop}%
\bibitem [{\citenamefont {P{\'e}rez}\ \emph {et~al.}(2015)\citenamefont
  {P{\'e}rez}, \citenamefont {Banerjee}, \citenamefont {Biraben}, \citenamefont
  {{Brook-Roberge}}, \citenamefont {Charlton}, \citenamefont {Clad{\'e}},
  \citenamefont {Comini}, \citenamefont {Crivelli}, \citenamefont {Dalkarov},
  \citenamefont {Debu}, \citenamefont {Douillet}, \citenamefont {Dufour},
  \citenamefont {Dupr{\'e}}, \citenamefont {Eriksson}, \citenamefont
  {Froelich}, \citenamefont {Grandemange}, \citenamefont {Guellati},
  \citenamefont {Gu{\'e}rout}, \citenamefont {Heinrich}, \citenamefont
  {Hervieux}, \citenamefont {Hilico}, \citenamefont {Husson}, \citenamefont
  {Indelicato}, \citenamefont {Jonsell}, \citenamefont {Karr}, \citenamefont
  {Khabarova}, \citenamefont {Kolachevsky}, \citenamefont {Kuroda},
  \citenamefont {Lambrecht}, \citenamefont {Leite}, \citenamefont {Liszkay},
  \citenamefont {Lunney}, \citenamefont {Madsen}, \citenamefont {Manfredi},
  \citenamefont {Mansouli{\'e}}, \citenamefont {Matsuda}, \citenamefont
  {Mohri}, \citenamefont {Mortensen}, \citenamefont {Nagashima}, \citenamefont
  {Nesvizhevsky}, \citenamefont {Nez}, \citenamefont {Regenfus}, \citenamefont
  {Rey}, \citenamefont {Reymond}, \citenamefont {Reynaud}, \citenamefont
  {Rubbia}, \citenamefont {Sacquin}, \citenamefont {{Schmidt-Kaler}},
  \citenamefont {Sillitoe}, \citenamefont {Staszczak}, \citenamefont
  {{Szabo-Foster}}, \citenamefont {Torii}, \citenamefont {Vallage},
  \citenamefont {Valdes}, \citenamefont {{Van der Werf}}, \citenamefont
  {Voronin}, \citenamefont {Walz}, \citenamefont {Wolf}, \citenamefont
  {Wronka},\ and\ \citenamefont {Yamazaki}}]{Perez2015}%
  \BibitemOpen
  \bibfield  {author} {\bibinfo {author} {\bibfnamefont {P.}~\bibnamefont
  {P{\'e}rez}}, \bibinfo {author} {\bibfnamefont {D.}~\bibnamefont {Banerjee}},
  \bibinfo {author} {\bibfnamefont {F.}~\bibnamefont {Biraben}}, \bibinfo
  {author} {\bibfnamefont {D.}~\bibnamefont {{Brook-Roberge}}}, \bibinfo
  {author} {\bibfnamefont {M.}~\bibnamefont {Charlton}}, \bibinfo {author}
  {\bibfnamefont {P.}~\bibnamefont {Clad{\'e}}}, \bibinfo {author}
  {\bibfnamefont {P.}~\bibnamefont {Comini}}, \bibinfo {author} {\bibfnamefont
  {P.}~\bibnamefont {Crivelli}}, \bibinfo {author} {\bibfnamefont
  {O.}~\bibnamefont {Dalkarov}}, \bibinfo {author} {\bibfnamefont
  {P.}~\bibnamefont {Debu}}, \bibinfo {author} {\bibfnamefont {A.}~\bibnamefont
  {Douillet}}, \bibinfo {author} {\bibfnamefont {G.}~\bibnamefont {Dufour}},
  \bibinfo {author} {\bibfnamefont {P.}~\bibnamefont {Dupr{\'e}}}, \bibinfo
  {author} {\bibfnamefont {S.}~\bibnamefont {Eriksson}}, \bibinfo {author}
  {\bibfnamefont {P.}~\bibnamefont {Froelich}}, \bibinfo {author}
  {\bibfnamefont {P.}~\bibnamefont {Grandemange}}, \bibinfo {author}
  {\bibfnamefont {S.}~\bibnamefont {Guellati}}, \bibinfo {author}
  {\bibfnamefont {R.}~\bibnamefont {Gu{\'e}rout}}, \bibinfo {author}
  {\bibfnamefont {J.~M.}\ \bibnamefont {Heinrich}}, \bibinfo {author}
  {\bibfnamefont {P.-A.}\ \bibnamefont {Hervieux}}, \bibinfo {author}
  {\bibfnamefont {L.}~\bibnamefont {Hilico}}, \bibinfo {author} {\bibfnamefont
  {A.}~\bibnamefont {Husson}}, \bibinfo {author} {\bibfnamefont
  {P.}~\bibnamefont {Indelicato}}, \bibinfo {author} {\bibfnamefont
  {S.}~\bibnamefont {Jonsell}}, \bibinfo {author} {\bibfnamefont {J.-P.}\
  \bibnamefont {Karr}}, \bibinfo {author} {\bibfnamefont {K.}~\bibnamefont
  {Khabarova}}, \bibinfo {author} {\bibfnamefont {N.}~\bibnamefont
  {Kolachevsky}}, \bibinfo {author} {\bibfnamefont {N.}~\bibnamefont {Kuroda}},
  \bibinfo {author} {\bibfnamefont {A.}~\bibnamefont {Lambrecht}}, \bibinfo
  {author} {\bibfnamefont {A.~M.~M.}\ \bibnamefont {Leite}}, \bibinfo {author}
  {\bibfnamefont {L.}~\bibnamefont {Liszkay}}, \bibinfo {author} {\bibfnamefont
  {D.}~\bibnamefont {Lunney}}, \bibinfo {author} {\bibfnamefont
  {N.}~\bibnamefont {Madsen}}, \bibinfo {author} {\bibfnamefont
  {G.}~\bibnamefont {Manfredi}}, \bibinfo {author} {\bibfnamefont
  {B.}~\bibnamefont {Mansouli{\'e}}}, \bibinfo {author} {\bibfnamefont
  {Y.}~\bibnamefont {Matsuda}}, \bibinfo {author} {\bibfnamefont
  {A.}~\bibnamefont {Mohri}}, \bibinfo {author} {\bibfnamefont
  {T.}~\bibnamefont {Mortensen}}, \bibinfo {author} {\bibfnamefont
  {Y.}~\bibnamefont {Nagashima}}, \bibinfo {author} {\bibfnamefont
  {V.}~\bibnamefont {Nesvizhevsky}}, \bibinfo {author} {\bibfnamefont
  {F.}~\bibnamefont {Nez}}, \bibinfo {author} {\bibfnamefont {C.}~\bibnamefont
  {Regenfus}}, \bibinfo {author} {\bibfnamefont {J.-M.}\ \bibnamefont {Rey}},
  \bibinfo {author} {\bibfnamefont {J.-M.}\ \bibnamefont {Reymond}}, \bibinfo
  {author} {\bibfnamefont {S.}~\bibnamefont {Reynaud}}, \bibinfo {author}
  {\bibfnamefont {A.}~\bibnamefont {Rubbia}}, \bibinfo {author} {\bibfnamefont
  {Y.}~\bibnamefont {Sacquin}}, \bibinfo {author} {\bibfnamefont
  {F.}~\bibnamefont {{Schmidt-Kaler}}}, \bibinfo {author} {\bibfnamefont
  {N.}~\bibnamefont {Sillitoe}}, \bibinfo {author} {\bibfnamefont
  {M.}~\bibnamefont {Staszczak}}, \bibinfo {author} {\bibfnamefont {C.~I.}\
  \bibnamefont {{Szabo-Foster}}}, \bibinfo {author} {\bibfnamefont
  {H.}~\bibnamefont {Torii}}, \bibinfo {author} {\bibfnamefont
  {B.}~\bibnamefont {Vallage}}, \bibinfo {author} {\bibfnamefont
  {M.}~\bibnamefont {Valdes}}, \bibinfo {author} {\bibfnamefont {D.~P.}\
  \bibnamefont {{Van der Werf}}}, \bibinfo {author} {\bibfnamefont
  {A.}~\bibnamefont {Voronin}}, \bibinfo {author} {\bibfnamefont
  {J.}~\bibnamefont {Walz}}, \bibinfo {author} {\bibfnamefont {S.}~\bibnamefont
  {Wolf}}, \bibinfo {author} {\bibfnamefont {S.}~\bibnamefont {Wronka}},\ and\
  \bibinfo {author} {\bibfnamefont {Y.}~\bibnamefont {Yamazaki}},\ }\bibfield
  {title} {\bibinfo {title} {The {{GBAR}} antimatter gravity experiment},\
  }\href {https://doi.org/10.1007/s10751-015-1154-8} {\bibfield  {journal}
  {\bibinfo  {journal} {Hyperfine Interactions}\ }\textbf {\bibinfo {volume}
  {233}},\ \bibinfo {pages} {21} (\bibinfo {year} {2015})}\BibitemShut
  {NoStop}%
\bibitem [{\citenamefont {Wagner}\ \emph {et~al.}(2012)\citenamefont {Wagner},
  \citenamefont {Schlamminger}, \citenamefont {Gundlach},\ and\ \citenamefont
  {Adelberger}}]{Wagner2012}%
  \BibitemOpen
  \bibfield  {author} {\bibinfo {author} {\bibfnamefont {T.~A.}\ \bibnamefont
  {Wagner}}, \bibinfo {author} {\bibfnamefont {S.}~\bibnamefont
  {Schlamminger}}, \bibinfo {author} {\bibfnamefont {J.~H.}\ \bibnamefont
  {Gundlach}},\ and\ \bibinfo {author} {\bibfnamefont {E.~G.}\ \bibnamefont
  {Adelberger}},\ }\bibfield  {title} {\bibinfo {title} {Torsion-balance tests
  of the weak equivalence principle},\ }\href
  {https://doi.org/10.1088/0264-9381/29/18/184002} {\bibfield  {journal}
  {\bibinfo  {journal} {Classical and Quantum Gravity}\ }\textbf {\bibinfo
  {volume} {29}},\ \bibinfo {pages} {184002} (\bibinfo {year}
  {2012})}\BibitemShut {NoStop}%
\bibitem [{\citenamefont {Touboul}\ and\ \citenamefont
  {al.}(2017)}]{Touboul2017}%
  \BibitemOpen
  \bibfield  {author} {\bibinfo {author} {\bibfnamefont {P.}~\bibnamefont
  {Touboul}}\ and\ \bibinfo {author} {\bibnamefont {al.}},\ }\bibfield  {title}
  {\bibinfo {title} {{MICROSCOPE Mission: First Results of a Space Test of the
  Equivalence Principle}},\ }\href
  {https://doi.org/10.1103/PhysRevLett.119.231101} {\bibfield  {journal}
  {\bibinfo  {journal} {Phys. Rev. Lett.}\ }\textbf {\bibinfo {volume} {119}},\
  \bibinfo {pages} {231101} (\bibinfo {year} {2017})}\BibitemShut {NoStop}%
\bibitem [{\citenamefont {Will}(2018)}]{Will2018}%
  \BibitemOpen
  \bibfield  {author} {\bibinfo {author} {\bibfnamefont {C.~M.}\ \bibnamefont
  {Will}},\ }\href@noop {} {\emph {\bibinfo {title} {{Theory and Experiment in
  Gravitational Physics (new edition)}}}}\ (\bibinfo  {publisher} {{Cambridge
  University Press}},\ \bibinfo {year} {2018})\BibitemShut {NoStop}%
\bibitem [{\citenamefont {Viswanathan}\ \emph {et~al.}(2018)\citenamefont
  {Viswanathan}, \citenamefont {Fienga}, \citenamefont {Minazzoli},
  \citenamefont {Bernus}, \citenamefont {Laskar},\ and\ \citenamefont
  {Gastineau}}]{Viswanathan2018}%
  \BibitemOpen
  \bibfield  {author} {\bibinfo {author} {\bibfnamefont {V.}~\bibnamefont
  {Viswanathan}}, \bibinfo {author} {\bibfnamefont {A.}~\bibnamefont {Fienga}},
  \bibinfo {author} {\bibfnamefont {O.}~\bibnamefont {Minazzoli}}, \bibinfo
  {author} {\bibfnamefont {L.}~\bibnamefont {Bernus}}, \bibinfo {author}
  {\bibfnamefont {J.}~\bibnamefont {Laskar}},\ and\ \bibinfo {author}
  {\bibfnamefont {M.}~\bibnamefont {Gastineau}},\ }\bibfield  {title} {\bibinfo
  {title} {{The new lunar ephemeris INPOP17a and its application to fundamental
  physics}},\ }\href {https://doi.org/10.1093/mnras/sty096} {\bibfield
  {journal} {\bibinfo  {journal} {Monthly Notices of the Royal Astronomical
  Society}\ }\textbf {\bibinfo {volume} {476}},\ \bibinfo {pages} {1877}
  (\bibinfo {year} {2018})}\BibitemShut {NoStop}%
\bibitem [{\citenamefont {Asenbaum}\ \emph {et~al.}(2020)\citenamefont
  {Asenbaum}, \citenamefont {Overstreet}, \citenamefont {Kim}, \citenamefont
  {Curti},\ and\ \citenamefont {Kasevich}}]{Asenbaum2020}%
  \BibitemOpen
  \bibfield  {author} {\bibinfo {author} {\bibfnamefont {P.}~\bibnamefont
  {Asenbaum}}, \bibinfo {author} {\bibfnamefont {C.}~\bibnamefont
  {Overstreet}}, \bibinfo {author} {\bibfnamefont {M.}~\bibnamefont {Kim}},
  \bibinfo {author} {\bibfnamefont {J.}~\bibnamefont {Curti}},\ and\ \bibinfo
  {author} {\bibfnamefont {M.~A.}\ \bibnamefont {Kasevich}},\ }\bibfield
  {title} {\bibinfo {title} {{Atom-Interferometric Test of the Equivalence
  Principle at the ${10}^{\ensuremath{-}12}$ Level}},\ }\href
  {https://doi.org/10.1103/PhysRevLett.125.191101} {\bibfield  {journal}
  {\bibinfo  {journal} {Phys. Rev. Lett.}\ }\textbf {\bibinfo {volume} {125}},\
  \bibinfo {pages} {191101} (\bibinfo {year} {2020})}\BibitemShut {NoStop}%
\bibitem [{\citenamefont {Walz}\ and\ \citenamefont
  {H{\"a}nsch}(2004)}]{Walz2004}%
  \BibitemOpen
  \bibfield  {author} {\bibinfo {author} {\bibfnamefont {J.}~\bibnamefont
  {Walz}}\ and\ \bibinfo {author} {\bibfnamefont {T.~W.}\ \bibnamefont
  {H{\"a}nsch}},\ }\bibfield  {title} {\bibinfo {title} {{A Proposal to Measure
  Antimatter Gravity Using Ultracold Antihydrogen Atoms}},\ }\href
  {https://doi.org/10.1023/B:GERG.0000010730.93408.87} {\bibfield  {journal}
  {\bibinfo  {journal} {General Relativity and Gravitation}\ }\textbf {\bibinfo
  {volume} {36}},\ \bibinfo {pages} {561} (\bibinfo {year} {2004})}\BibitemShut
  {NoStop}%
\bibitem [{\citenamefont {Rousselle}\ \emph {et~al.}(2021)\citenamefont
  {Rousselle}, \citenamefont {Clad\'e}, \citenamefont {Guellati-Khelifa},
  \citenamefont {Gu\'erout},\ and\ \citenamefont {Reynaud}}]{Rousselle2021}%
  \BibitemOpen
  \bibfield  {author} {\bibinfo {author} {\bibfnamefont {O.}~\bibnamefont
  {Rousselle}}, \bibinfo {author} {\bibfnamefont {P.}~\bibnamefont {Clad\'e}},
  \bibinfo {author} {\bibfnamefont {S.}~\bibnamefont {Guellati-Khelifa}},
  \bibinfo {author} {\bibfnamefont {R.}~\bibnamefont {Gu\'erout}},\ and\
  \bibinfo {author} {\bibfnamefont {S.}~\bibnamefont {Reynaud}},\ }\bibfield
  {title} {\bibinfo {title} {Analysis of the timing of freely falling
  antihydrogen},\ }\Eprint {https://arxiv.org/abs/2111.02815} {arXiv:2111.02815
  [physics.atom-ph]}  (\bibinfo {year} {2021}),\ \bibinfo {note}
  {(submitted)}\BibitemShut {NoStop}%
\bibitem [{\citenamefont {Hilico}\ \emph {et~al.}(2014)\citenamefont {Hilico},
  \citenamefont {Karr}, \citenamefont {Douillet}, \citenamefont {Indelicato},
  \citenamefont {Wolf},\ and\ \citenamefont {Schmidt-Kaler}}]{Hilico2014}%
  \BibitemOpen
  \bibfield  {author} {\bibinfo {author} {\bibfnamefont {L.}~\bibnamefont
  {Hilico}}, \bibinfo {author} {\bibfnamefont {J.-P.}\ \bibnamefont {Karr}},
  \bibinfo {author} {\bibfnamefont {A.}~\bibnamefont {Douillet}}, \bibinfo
  {author} {\bibfnamefont {P.}~\bibnamefont {Indelicato}}, \bibinfo {author}
  {\bibfnamefont {S.}~\bibnamefont {Wolf}},\ and\ \bibinfo {author}
  {\bibfnamefont {F.}~\bibnamefont {Schmidt-Kaler}},\ }\bibfield  {title}
  {\bibinfo {title} {{Preparing single ultra-cold antihydrogen atoms for
  free-fall in GBAR}},\ }\href {https://doi.org/10.1142/S2010194514602695}
  {\bibfield  {journal} {\bibinfo  {journal} {International Journal of Modern
  Physics: Conference Series}\ }\textbf {\bibinfo {volume} {30}},\ \bibinfo
  {pages} {1460269} (\bibinfo {year} {2014})}\BibitemShut {NoStop}%
\bibitem [{\citenamefont {Sillitoe}\ \emph {et~al.}(2017)\citenamefont
  {Sillitoe}, \citenamefont {Karr}, \citenamefont {Heinrich}, \citenamefont
  {Louvradoux}, \citenamefont {Douillet},\ and\ \citenamefont
  {Hilico}}]{Sillitoe2017}%
  \BibitemOpen
  \bibfield  {author} {\bibinfo {author} {\bibfnamefont {N.}~\bibnamefont
  {Sillitoe}}, \bibinfo {author} {\bibfnamefont {J.-P.}\ \bibnamefont {Karr}},
  \bibinfo {author} {\bibfnamefont {J.}~\bibnamefont {Heinrich}}, \bibinfo
  {author} {\bibfnamefont {T.}~\bibnamefont {Louvradoux}}, \bibinfo {author}
  {\bibfnamefont {A.}~\bibnamefont {Douillet}},\ and\ \bibinfo {author}
  {\bibfnamefont {L.}~\bibnamefont {Hilico}},\ }\bibinfo {title}
  {{$\bar{\text{H}}^{+}$ Sympathetic Cooling Simulations with a Variable Time
  Step}},\ in\ \href {https://doi.org/10.7566/JPSCP.18.011014} {\emph {\bibinfo
  {booktitle} {Proceedings of the 12th International Conference on Low Energy
  Antiproton Physics (LEAP2016)}}},\ \bibinfo {series} {JPS Conf. Proc.},
  Vol.~\bibinfo {volume} {18}\ (\bibinfo {year} {2017})\BibitemShut {NoStop}%
\bibitem [{\citenamefont {Dufour}\ \emph {et~al.}(2013)\citenamefont {Dufour},
  \citenamefont {G\'erardin}, \citenamefont {Gu\'erout}, \citenamefont
  {Lambrecht}, \citenamefont {Nesvizhevsky}, \citenamefont {Reynaud},\ and\
  \citenamefont {Voronin}}]{Dufour2013}%
  \BibitemOpen
  \bibfield  {author} {\bibinfo {author} {\bibfnamefont {G.}~\bibnamefont
  {Dufour}}, \bibinfo {author} {\bibfnamefont {A.}~\bibnamefont {G\'erardin}},
  \bibinfo {author} {\bibfnamefont {R.}~\bibnamefont {Gu\'erout}}, \bibinfo
  {author} {\bibfnamefont {A.}~\bibnamefont {Lambrecht}}, \bibinfo {author}
  {\bibfnamefont {V.~V.}\ \bibnamefont {Nesvizhevsky}}, \bibinfo {author}
  {\bibfnamefont {S.}~\bibnamefont {Reynaud}},\ and\ \bibinfo {author}
  {\bibfnamefont {A.~Y.}\ \bibnamefont {Voronin}},\ }\bibfield  {title}
  {\bibinfo {title} {{Quantum reflection of antihydrogen from the Casimir
  potential above matter slabs}},\ }\href
  {https://doi.org/10.1103/PhysRevA.87.012901} {\bibfield  {journal} {\bibinfo
  {journal} {Phys. Rev. A}\ }\textbf {\bibinfo {volume} {87}},\ \bibinfo
  {pages} {012901} (\bibinfo {year} {2013})}\BibitemShut {NoStop}%
\bibitem [{\citenamefont {Fr\'echet}(1943)}]{Frechet}%
  \BibitemOpen
  \bibfield  {author} {\bibinfo {author} {\bibfnamefont {M.}~\bibnamefont
  {Fr\'echet}},\ }\bibfield  {title} {\bibinfo {title} {Sur l'extension de
  certaines \'evaluations statistiques au cas de petits \'echantillons},\
  }\href {https://doi.org/10.2307/1401114} {\bibfield  {journal} {\bibinfo
  {journal} {Review of the International Statistical Institute}\ }\textbf
  {\bibinfo {volume} {11}},\ \bibinfo {pages} {182} (\bibinfo {year}
  {1943})}\BibitemShut {NoStop}%
\bibitem [{\citenamefont {Cram\'er}(1999)}]{Cramer}%
  \BibitemOpen
  \bibfield  {author} {\bibinfo {author} {\bibfnamefont {H.}~\bibnamefont
  {Cram\'er}},\ }\href@noop {} {\emph {\bibinfo {title} {Mathematical Methods
  of Statistics (new edition)}}}\ (\bibinfo  {publisher} {{Princeton University
  Press}},\ \bibinfo {year} {1999})\BibitemShut {NoStop}%
\bibitem [{\citenamefont {R{\'e}fr{\'e}gier}(2004)}]{Refregier}%
  \BibitemOpen
  \bibfield  {author} {\bibinfo {author} {\bibfnamefont {P.}~\bibnamefont
  {R{\'e}fr{\'e}gier}},\ }\href@noop {} {\emph {\bibinfo {title} {Noise Theory
  and Application to Physics: From Fluctuations to Information}}},\ Advanced
  Texts in Physics\ (\bibinfo  {publisher} {{Springer}},\ \bibinfo {address}
  {{New York}},\ \bibinfo {year} {2004})\BibitemShut {NoStop}%
\bibitem [{\citenamefont {Dufour}(2015)}]{Dufour2015}%
  \BibitemOpen
  \bibfield  {author} {\bibinfo {author} {\bibfnamefont {G.}~\bibnamefont
  {Dufour}},\ }\bibfield  {title} {\bibinfo {title} {Quantum reflection from
  the {Casimir-Polder} potential},\ }\href@noop {} {\bibfield  {journal}
  {\bibinfo  {journal} {Thesis Laboratoire Kastler Brossel}\ } (\bibinfo {year}
  {2015})}\BibitemShut {NoStop}%
\bibitem [{\citenamefont {Dufour}\ \emph {et~al.}(2014)\citenamefont {Dufour},
  \citenamefont {Debu}, \citenamefont {Lambrecht}, \citenamefont
  {Nesvizhevsky}, \citenamefont {Reynaud},\ and\ \citenamefont
  {Voronin}}]{Dufour2014}%
  \BibitemOpen
  \bibfield  {author} {\bibinfo {author} {\bibfnamefont {G.}~\bibnamefont
  {Dufour}}, \bibinfo {author} {\bibfnamefont {P.}~\bibnamefont {Debu}},
  \bibinfo {author} {\bibfnamefont {A.}~\bibnamefont {Lambrecht}}, \bibinfo
  {author} {\bibfnamefont {V.~V.}\ \bibnamefont {Nesvizhevsky}}, \bibinfo
  {author} {\bibfnamefont {S.}~\bibnamefont {Reynaud}},\ and\ \bibinfo {author}
  {\bibfnamefont {A.~Y.}\ \bibnamefont {Voronin}},\ }\bibfield  {title}
  {\bibinfo {title} {Shaping the distribution of vertical velocities of
  antihydrogen in {GBAR}},\ }\href
  {https://doi.org/10.1140/epjc/s10052-014-2731-8} {\bibfield  {journal}
  {\bibinfo  {journal} {The European Physical Journal C}\ }\textbf {\bibinfo
  {volume} {74}},\ \bibinfo {pages} {2731} (\bibinfo {year}
  {2014})}\BibitemShut {NoStop}%
\bibitem [{\citenamefont {CERN}()}]{GBARpage}%
  \BibitemOpen
  \bibfield  {author} {\bibinfo {author} {\bibnamefont {CERN}},\ }\href@noop {}
  {\bibinfo {title} {{{GBAR}} collaboration}},\ \bibinfo {howpublished}
  {https://gbar.web.cern.ch/}\BibitemShut {NoStop}%
\end{thebibliography}%

\end{document}